%% file: main.tex
\documentclass[trackchanges,preprint2]{aastex631}

\usepackage{soul}
\usepackage{natbib}
\usepackage{amsmath}
\usepackage{gensymb}
\graphicspath{{./}{}}

\usepackage{color}

\usepackage{ulem}

\revised{\today}
\submitjournal{ApJ}

\shorttitle{Twelve-Year Cosmic Ray Anisotropy}
\shortauthors{IceCube Collaboration}

\begin{document}

\title{Observation of Cosmic-Ray Anisotropy in the Southern Hemisphere with 12 yr of Data Collected by the IceCube Neutrino Observatory}

\include{authors_i3}




\begin{abstract}
We analyzed the 7.92$ \times 10^{11}$ cosmic-ray-induced muon events collected by the IceCube Neutrino Observatory from May 13, 2011, when the fully constructed experiment started to take data, to May 12, 2023. This dataset provides an up-to-date cosmic-ray arrival direction distribution in the Southern Hemisphere with unprecedented statistical accuracy covering more than a full period length of a solar cycle. Improvements in Monte Carlo event simulation and better handling of year-to-year differences in data processing significantly reduce systematic uncertainties below the level of statistical fluctuations compared to the previously published results. We confirm the observation of a change in the angular structure of the cosmic-ray anisotropy between 10\,TeV and 1\,PeV, more specifically in the 100--300\,TeV energy range. 
For the first time, we analyzed the angular power spectrum at different energies. The observed variations of the power spectra with energy suggest relatively reduced large-scale features at high energy compared to those of medium and small scales.
%
The large volume of data enhances the statistical significance at higher energies, up to the PeV scale, and smaller angular scales, down to approximately 6$^{\circ}$ compared to previous findings.
\end{abstract}

\keywords{astroparticle physics, cosmic rays, ISM: magnetic fields}





\section{Introduction}
\label{sec:intro}
It is widely believed that most cosmic-ray particles detected on Earth below EeV energy are produced within the Milky Way galaxy. Their energy spectrum and arrival directions are determined by the properties and distribution of their galactic sources and by their propagation through the turbulent and inhomogeneous interstellar magnetized plasma. Although cosmic-ray properties have been investigated for over a century, it was only in the last thirty years that extensive ground-based experiments have accumulated high-quality observations of the arrival direction distribution of cosmic rays~\citep{nagashima_1998, hall_1999, guillian_2007, Aglietta_2009, Abdo_2009, munakata_2010, MINOS:2011icrc, aartsen2016, Amenomori:2017jbv, Bartoli_2018, Abeysekara_2018, Abeysekara_2019, Apel_2019, Ahlers_2019, He:202379}.

Recent observations (see references in~\cite{Ahlers:2016rox}) reveal a complex anisotropy in cosmic-ray arrival directions thanks to a large volume of accumulated data, the stability of experiments' instrumentation, and improved data analysis techniques. In particular, they show that the observed anisotropy is at the level of 10$^{-4}$--10$^{-3}$ in the TeV-PeV energy range. We can describe the observation as an intricate superposition of angular structures typically expressed with spherical harmonic functions. This mathematical description makes it possible to investigate the origin of the observations. For instance, it allows us to inspect the arrival direction distribution in relation to particle diffusion~\citep{Mertsch:2015jan, schlickeiser2019anisotropy, zhang2022cosmic, reichherzer2022anisotropic} and features in the local interstellar medium (ISM), including the heliosphere~\citep{Lazarian:2010sq, Desiati:2011xg, Schwadron2014, Zhang:2014jul, 2016JPhCS.767a2027Z, 0004-637X-842-1-54}. Moreover, we can determine the angular power spectrum, which quantitatively describes each angular scale's contributions to the observed distribution in terms of particle propagation through the turbulent magnetized plasma~\citep{blasi2004small, barquero2016, harding2016explaining, Ahlers:2016rox, kuhlen2021cosmic}. 

An intriguing observation provided by several experiments (see the reference list in the first paragraph of this section) is the evolution of the cosmic-ray arrival direction distribution as a function of energy. The global relative excess appears to shift shape and direction across the sky between 100\,TeV and 300\,TeV while staying relatively stable at lower and higher energies.
This change may indicate an influence of local sources or significant changes at larger distances within our local ISM.

In this paper, we report the cosmic-ray anisotropy results obtained by the IceCube Neutrino Observatory, with 7.92$ \times 10^{11}$ cosmic-ray events collected over a period of twelve years. In Sec.~\ref{sec:icecube}, we describe the IceCube experiment, including the data used in this analysis and the simulation used to estimate the cosmic-ray energy of each event in the sample. In Sec.~\ref{sec:analysis}, we describe how the analysis is performed, what methods and tools we used, and the results. We then describe how we estimated the systematic uncertainties associated with the anisotropy measurement in Sec.~\ref{sec:syst} and conclude with a summary discussion in Sec.~\ref{sec:summary} along with an outlook in Sec.~\ref{sec:outlook} on how this observation can contribute to our understanding of cosmic-ray physics and what additional measurements are necessary.

\section{The IceCube Experiment and Data} \label{sec:icecube}
This section describes the IceCube Neutrino Observatory and its various components (Sec.~\ref{ssec:detector}). We then summarize the nature of the experimental data (Sec.~\ref{ssec:exp}) as well as the production and use of simulation data (Sec.~\ref{ssec:sim}). 

\subsection{Detector} \label{ssec:detector}
The IceCube Neutrino Observatory~\citep{Aartsen_2017} is a cubic-kilometer-scale detector located at the geographic South Pole. The largest component of the detector is an \textit{in-ice} array buried deep in the Antarctic ice shelf, consisting of 5,160 digital optical modules (DOMs), extending from a depth of 1,450 meters to about 2,450 meters. The in-ice array is augmented with a denser subarray called DeepCore~\citep{DeepCore}. IceCube  also includes a \textit{surface} array called IceTop~\citep{IceCube:2013feb}, composed of tanks with DOMs inserted in transparent ice and designed to detect cosmic-ray showers above 100\,TeV. This paper describes an investigation performed with the high-energy cosmic-ray muon data collected by the IceCube in-ice array.

IceCube's DOMs consist of a 10-inch photomultiplier tube (PMT) and associated electronics for detection, digitization, and readout, all enclosed in a pressure-protective glass sphere. The in-ice array features DOMs arranged on 86 vertical \textit{strings} separated by an average distance of 125\,m, with 60 DOMs spaced every 17\,m on each string. The strings that comprise the DeepCore subarray in the center of IceCube are arranged with a horizontal separation of about 70\,m and a vertical DOM spacing of 7\,m on each string.

The IceCube in-ice array is designed to detect high-energy neutrinos of astrophysical origin. Neutrinos are observed through the secondary charged particles generated by their interaction in the ice inside and outside the instrumented volume or in the bedrock below. Such secondary particles, being relativistic, induce Cherenkov photons that propagate through the ice and are detected by the DOMs. The PMT produces a current pulse whenever a Cherenkov photon ejects an electron from its cathode; these pulses are digitized and time-stamped by the DOM onboard electronics before being sent to a data acquisition system for trigger determination and subsequent event 
reconstruction. To identify a physical event and filter out background signals caused by the 500-Hz PMTs' dark noise, neighboring DOMs on the same string must satisfy a local coincidence in time within $\pm 1\,\mu$s. A trigger is issued when eight or more DOMs satisfying the local coincidence criterion record photons within a $\pm 5\,\mu$s time window~\citep{ABBASI2009294, Aartsen_2017}. 

All events that trigger the IceCube detector are reconstructed using a likelihood-based method that accounts for light propagation in the ice. 
This Single-Photo-Electron (SPE) reconstruction maximizes the likelihood constructed from the probability density function for the arrival times of single photons at the locations of activated optical modules
to reconstruct the muon track~\citep{Ahrens2004169}. This SPE likelihood algorithm is seeded with the results of a fast \textit{linefit} reconstruction that minimizes the sum of the squares of the distances between the track and the recorded pulses at each optical module~\citep{AARTSEN2014143}.

Most of the events collected by the in-ice array are high-energy muons produced by cosmic-ray particles colliding with Earth’s atmosphere; the muons have a mean energy at the surface of the ice sheet of about 1 TeV. These events constitute a significant background for neutrino event identification and analysis. However, they also provide an opportunity to study cosmic rays with IceCube. At these energies, cosmic-ray muons form narrow bundles that 
share the direction of the parent cosmic-ray particle to within a fraction of a degree. Therefore, they can be used to measure the cosmic rays' arrival direction distribution at energies of about 10\,TeV and above. 
The directions of muon bundles are reconstructed with a single muon hypothesis.
Data from IceTop are not included in this analysis; a dedicated study of that dataset is underway (see Sec.~\ref{sec:outlook}).

\subsection{Experimental Data} \label{ssec:exp}
This analysis uses twelve full years of data collected by the completed IceCube detector (referred to as \textit{IC86}) from May 13, 2011, through May 12, 2023. The data correspond to the \textit{good run list} established by IceCube's online monitoring system~\citep{Aartsen_2017}.

The rate of events collected by the IceCube in-ice array used in this work is between 2.0 and 2.3\,kHz, following the yearly modulation of the stratospheric temperature over the Antarctic continent~\citep{Gaisser_2021,  VERPOEST2024102985}. The volume of data collected amounts to about 1\,TB per day. The limited bandwidth of the communication satellites that provide connectivity to the South Pole sets a practical limit on the amount of information that can be stored and transferred from each event that activates the detector's trigger. For this reason, we utilize a compact data storage and transfer (DST) format to store the reconstruction results of all recorded events performed in situ before transferring the data to the Northern Hemisphere.

The DST format stores the event time $t$, directional angles zenith and azimuth ($\theta, \varphi$) of the reconstructed track in the detector reference system, the number of activated modules, the number of photoelectrons recorded in the detector, and additional variables used for reconstruction quality selections. The data are encoded and compressed for a total of about 3\,GB per day.
The DST data sequence is the primary source of information from the IceCube in-ice array on cosmic-ray arrival directions for anisotropy analyses and was primarily designed for this purpose.

\begin{table*}[t!]
\begin{center}
\footnotesize
\setlength{\tabcolsep}{14pt}
\begin{tabular}{c|cccccc|c}
\hline 
Year           & 2011 & 2012 & 2013 & 2014 & 2015 & 2016 \\ \hline
Livetime[days] & 341.33
               & 343.88
               & 357.91
               & 357.68
               & 361.47
               & 359.89 \\
Duty Cycle     & 93.26\%
               & 94.21\%
               & 98.06\%
               & 97.99\%
               & 98.76\%
               & 98.60\% \\
Events ($\times 10^{10}$) & 6.09
                         & 6.39
                         & 6.66
                         & 6.66
                         & 6.66
                         & 6.67 \\ \hline \hline
Year           & 2017 & 2018 & 2019 & 2020 & 2021 & 2022 & \textbf{Total} \\ \hline
Livetime[days] & 359.31
               & 361.84
               & 361.62
               & 363.28
               & 364.25
               & 362.85
               & \textbf{4295.29} \\
Duty Cycle     & 98.44\%
               & 99.13\%
               & 98.80\%
               & 99.53\%
               & 99.80\%
               & 99.41\% 
               & \textbf{98.00}\% \\
Events ($\times 10^{10}$) & 6.65
                         & 6.67
                         & 6.73
                         & 6.67
                         & 6.67
                         & 6.67
                         & \textbf{79.18}\\ \hline
\end{tabular}
\end{center}
\caption{The good run livetime and the number of cosmic-ray muon events collected by IceCube from May 13 of each year through May 12 of the following year, starting from 2011 (after the detector's completion). Also shown is the corresponding duty cycle, which accounts for data runs deemed unstable and not appropriate for physics analyses.}
\label{table:data}
\end{table*}
During this twelve-year analysis period, IceCube accumulated a total livetime of 4295.29 days (98\% of the twelve full years), collecting 792 billion cosmic-ray muon events. Table~\ref{table:data} shows each calendar year's event counts and the corresponding good run livetime.

\subsection{Simulation Data} \label{ssec:sim}

Simulation data are used in this analysis to: 1) assess the directional angular resolution of the reconstructed events to define the field of view of the arrival direction distribution and 2) estimate event energy to design a selection dedicated to splitting the event sample into different cosmic-ray energy groups (see Sec.~\ref{sec:analysis}). To generate cosmic-ray muon events triggering the IceCube detector, we use CORSIKA~\citep{Heck:1998vt}, which is a physics software package for the simulation of extensive air showers induced by high-energy primary cosmic rays, such as protons and heavier nuclei. 
The cosmic-ray anisotropy study presented here does not use Monte Carlo simulation data to generate sky maps.

CORSIKA makes use of importance sampling Monte Carlo technique~\citep{2c3b6b76-0a89-36bb-8a21-b89506ce74ac}
to generate a cosmic-ray primary flux. We produce data with different fractions of five primary nuclei (H, He, N, Al, and Fe) to represent different groups of chemical elements with similar atomic masses, using an arbitrary power-law spectrum with index $\gamma_{_{_Z}} > 1.0$ given by
\begin{equation}\label{eq:flux}
    d\Phi_Z/dE=\Phi_Z^0 E^{-\gamma_{_{_Z}}} \; . 
\end{equation}
The relative fractions $\Phi_Z^0$ of each of the five elements are based on the cosmic-ray flux from the Gaisser H3a model~\citep{Gaisser:2012zz} at 1\,TeV. We generate an energy spectrum of $\gamma_{_{_Z}} = 2.6$, roughly approximating the H3a model. The simulated five-component CORSIKA samples are then weighted to the H3a model to reproduce the natural energy spectrum and composition. 

CORSIKA simulates particle interactions and decays in the atmosphere. Hadronic interactions are computed using the SIBYLL-2.3c mini-jet high-energy hadronic model~\citep{PhysRevD.100.103018} at energies above 80\,GeV. Electromagnetic interactions are calculated using the EGS4 model~\citep{Nelson:1985ec}. In all cases, we do not simulate particles below 270\,GeV as they cannot reach the in-ice array.
To simulate particle propagation through the atmosphere, CORSIKA implements several atmospheric models, including four South Pole atmosphere profiles parameterized according to the MSIS-90-E model~\citep{msis-e-90} for March 31, July 01, Oct.~01, and Dec.~31 of 1997. IceCube Monte Carlo production generates equal amounts of the four MSIS-90-E atmospheres. The final result of the CORSIKA simulation is muon bundles recorded at the ice sheet surface.

We propagate the muons generated by CORSIKA through the ice and simulate the in-ice array response. The detector simulation --- including detailed photon propagation, PMT response, and acquisition electronics --- uses IceCube's IceTray framework~\citep{DeYoungICETRAYA} with an interface to read CORSIKA formatted files. 
This simulation uses only high-energy muons reaching the in-ice array; all other shower components not contributing to the physical event are omitted to reduce memory requirements and computation time.
Finally, the triggered muon events are reconstructed and processed with the same pipeline used for experimental data. .

\section{Analysis \& Results} \label{sec:analysis}

In this section, we describe the methods employed for calculating the arrival-direction-distribution sky maps (Sec.~\ref{ssec:makingmaps}) and for splitting the experimental data into different cosmic-ray energy bins (Sec.~\ref{ssec:makingene}). We present the cosmic-ray anisotropy sky maps in equatorial coordinates as a function of energy in Sec.~\ref{ssec:maps}. We compare the measurement of the amplitude and phase of the dipole component with other experiments in Sec.~\ref{ssec:dipole}. We also report on the angular structure, including the angular power spectrum in Sec.~\ref{ssec:ang} and the corresponding small-scale relative intensity sky maps in Sec.~\ref{ssec:small-scale}.

\subsection{Constructing Arrival Direction Sky Maps} \label{ssec:makingmaps}

Numerous methods are used to measure the anisotropies in cosmic-ray arrival direction, depending on the experiment being used, including: Rayleigh amplitude method~\citep{PhysRevLett.34.1530, Antoni_2004}, time-scrambling~\citep{Alexandreas1993}, direct integration~\citep{Milagro:2003yym}, a $\chi^2$ iterative method optimized for gamma-ray searches~\citep{Amenomori_2005}, East-West~\citep{Bonino_2011}, and the iterative maximum likelihood method developed by~\cite{0004-637X-823-1-10}. This last method 
was utilized in the combined analysis with HAWC in~\cite{Abeysekara_2019}. It is designed to correctly recover the amplitude of large-scale features in the anisotropy that are attenuated as a result of the difference between the instantaneous and integrated field of view (FoV). For detectors located away from the poles, the instantaneous exposure varies during a 24 hr period and does not match the full daily exposure. However, as noted by~\cite{0004-637X-823-1-10}, since IceCube's instantaneous FoV is identical to the time-integrated one, the iterative method is equivalent to direct integration or time scrambling.
For this reason, in this analysis, we employed the time-scrambling method, as in previous IceCube analyses~\citep{IceCube:2012feb, aartsen2016}, to provide a ``reference sky map" representing the experiment's response to an isotropic cosmic-ray flux. This reference sky map also accounts for possible detection biases (e.g., nonuniform exposure and gaps in the experiment's livetime). 

The sky maps are created and analyzed using the Hierarchical Equal Area isoLatitude Pixelation, \texttt{HEALPix}\footnote{http://healpix.sourceforge.net}. Originally used to store and study cosmic microwave background data~\citep{Gorski_2005}, HEALPix is a data structure with a library of computational algorithms and visualization software. It enables fast scientific applications on large volumes of astronomical data and surveys in spherical maps subdivided into equal-area pixels. For this work, we use pixels with a size of $(0.84^{\circ})^2$, obtained using the HEALPix parameter $N_{\rm side}=64$. With this choice, the pixel size is smaller than the statistical uncertainty of the muon directional reconstruction. for this work we utilize \texttt{healpy}~\citep{Zonca2019}, the Python~\citep{10.5555/1593511} implementation of \texttt{HEALPix}.

The time-scrambling method is a data-driven approach that uses the number of reconstructed events per pixel to determine the reference sky map. We create a new dataset of events by keeping the local coordinates unchanged but assigning times randomly chosen from events in the original dataset with a time window $\Delta t$.
The time window can be flexibly selected, depending on the angular scale one is searching for; we use a window of $\Delta t= 24$\,h, providing sensitivity to all angular scales. For IceCube, the time reassignment (while keeping the events' local coordinates constant) scrambles/shuffles the event's right ascension (RA) but not its declination. To reduce the statistical fluctuations in the reference sky map, we produce 20 independently scrambled events for every event in the original data, adding each with a weight of 1/20.

After creating the reference sky maps,
the cosmic ray anisotropy is quantified by calculating the relative intensity, which is defined as
\begin{equation}\label{eq:relint}
    \delta I_i = \frac{N_i- \langle N \rangle_i}{\langle N \rangle_i} \; ,
\end{equation}
where $N_i$ and $\langle N \rangle_i$ represent the number of events for pixel $i$ in the data and reference sky maps, respectively. The statistical significance of each pixel is estimated using the method developed by~\cite{LiMa:1983sep}. 
We smooth the data and the reference sky maps for visualization purposes to observe large- and small-scale anisotropies at high significance (see Sec.~\ref{ssec:maps}). We use a process known as \textit{top-hat} smoothing, which assigns a value in each pixel equal to the sum of all pixels' content within a smoothing radius. Our default smoothing radius is $20^\circ$, consistent with previous IceCube publications~\citep{IceCube:2011oct,IceCube:2012feb, aartsen2016}. Analyses of the cosmic-ray arrival direction distributions are conducted on unsmoothed data and reference maps.

The Monte Carlo simulation data described in Sec.~\ref{ssec:sim} indicate that the median angular resolution of the entire data sample is about $3^\circ$ (reaching $1^\circ$ above 100 TeV). However, it surpasses $20^\circ$ above a zenith angle of $50^\circ$ at 10 TeV, with a significant improvement at higher energy.
The poor angular resolution at large zenith angles results from the increasing energy threshold of cosmic-ray primaries, as muons must travel increasingly longer distances through the ice. As a result, the number of hit optical sensors decreases at larger zenith angles for a given primary energy~\citep{IceCube:2012feb,Abeysekara_2019}.
For visualization purposes, sky maps are masked to omit events with reconstructed zenith angles above $65^\circ$, above which the event rate is negligible. On the other hand, smaller sky coverage is used for quantitative data analyses (see Sec.~\ref{ssec:dipole}).
This study does not attempt to select events with better angular resolution, as we aim to obtain the largest data volume to determine anisotropy structures down to approximately $6^{\circ}$ (see Sec.~\ref{ssec:ang}).

\subsection{Constructing Energy Datasets} \label{ssec:makingene}

\begin{figure*}[t]
  \centering
  \includegraphics[width=0.95\textwidth]{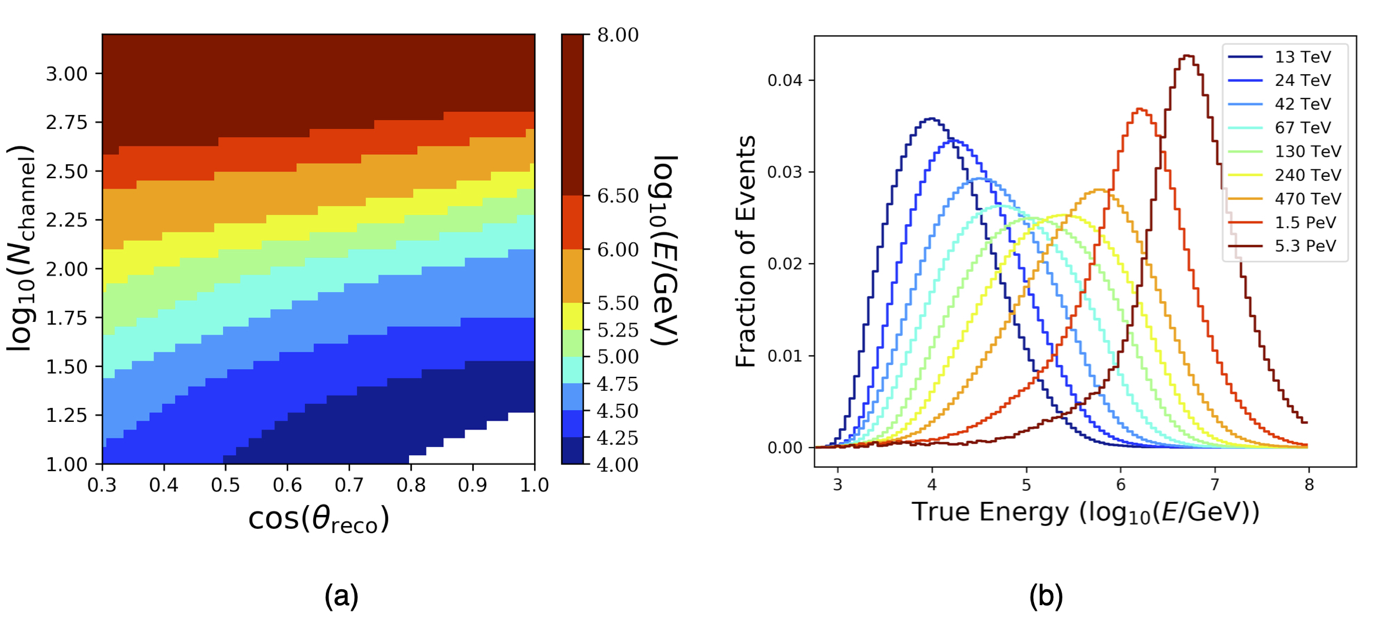}
     \caption{(a) Median true primary cosmic-ray energy for muon events, binned by the number of DOMs triggered $\log_{10}(N_\mathrm{channel})$ and the reconstructed zenith angle $\cos(\theta_\mathrm{reco})$ (with bin sizes of 0.044, 0.07, respectively). The nine color areas indicate the regions in parameter space used to split the data into the nine data sets with a given median primary energy. (b) The true energy distributions for each of the nine energy sets. The vertical scale indicates the fraction of events per energy bin of size $\Delta \log_{10}(E) = 0.05$ for each sample. The legend contains the median value for each distribution.}
     \label{fig:med_energy}
\end{figure*}

To study the energy dependence of the cosmic-ray anisotropy, we split the experimental data into several sets of increasing cosmic-ray energy. This procedure uses the simulation data described in Sec.~\ref{ssec:sim}.
In earlier studies on anisotropy, we relied on the data gathered during the construction of IceCube, which limited our capacity to generate sufficient simulation data for each experiment's configuration (see~\cite{aartsen2016} and references therein). However, we now benefit from exclusively using the final detector configuration for the entire time period used in this work. 

The primary cosmic ray energy estimation is done on a statistical basis using the number of DOMs triggered by each event ($N_\mathrm{channel}$) and the corresponding muon trajectory's reconstructed zenith angle ($\theta_\mathrm{reco}$). Higher deposited energy in the experiment triggers more DOMs. This happens with high-energy muons and high-multiplicity muon bundles, which are associated with more energetic or heavier cosmic-ray primary particles. In addition, events with a larger zenith angle must penetrate more ice before passing through IceCube's in-ice array. Inclined muons therefore have higher mean energy at the surface, corresponding to higher primary cosmic-ray energy~\citep{AARTSEN20161}. We use the simulation data described in Sec.~\ref{ssec:sim} to determine the dependency between the experiment's response and the cosmic-ray primary particle energy as a function of the reconstructed zenith angle.

Using the procedure described in~\cite{aartsen2016}, Monte Carlo simulated events were placed into a three-dimensional histogram with axes corresponding to the observed $\log_{10}(N_\mathrm{channel})$, $\cos(\theta_\mathrm{reco})$, and the logarithm of the true energy of the primary cosmic-ray particle (bin sizes of 0.044, 0.07, and 0.05, respectively). After the histogram is populated, we select the median true energy value for each bin, effectively transforming the histogram into a lookup table that displays median energy as a function of $\log_{10}(N_\mathrm{channel})$ and $\cos(\theta_\mathrm{reco})$. The table is smoothed with a B-spline function~\citep{WHITEHORN20132214} to avoid statistical artifacts resulting from the limited simulation at high energies. The end result is used to define nine ranges of median primary cosmic-ray energy, as shown in Fig.~\ref{fig:med_energy}a. The bands in the figure provide a proxy for estimating the primary energy of events, based on the measured variables $N_\mathrm{channel}$ and $\theta_\mathrm{reco}$. The described energy selection method yields an energy distribution in each event set defined by a colored band with a median value that is fairly constant as a function of $\theta_{reco}$. To ensure that the median energy on the lowest energy bin is also constant as a function of zenith angle, we exclude the events populating the white area in the bottom corner of Fig.~\ref{fig:med_energy}a.
Since the anisotropy changes with energy (see Sec.~\ref{ssec:maps}), ensuring a uniform energy response across the FoV provides an unbiased measurement of the arrival direction distribution.

\begin{table}[h!]
\begin{center}
    \footnotesize
    \begin{tabular}{c | c}
        \hline
        $E_\mathrm{med}$ [$\mathrm{TeV}$] & Events [$10^9$] \\
        \hline
        13 & 330.40 \\
        24 & 197.15 \\
        42 & 89.74 \\
        67 & 22.17 \\
        130 & 6.32 \\
        240 & 2.13 \\
        470 & 1.0147 \\
        1500 & 0.1019 \\
        5300 & 0.0128 \\
        \hline
    \end{tabular}
\end{center}
\caption{Total number of events in each of the nine energy sets with the corresponding median energy.} 
\label{table:enevents}
\end{table}
Table~\ref{table:enevents} displays the number of events selected for each energy set used in the analysis. This corresponds to about 82\% of the total number of events collected by the IceCube in-ice array (see Table~\ref{table:data}), a reduction stemming from the limited FoV (see Sec.~\ref{ssec:makingmaps}), the exclusion of the lower right corner ($\log_{10}(N_\mathrm{channel})$, $\cos(\theta_\mathrm{reco})$) region in Fig.~\ref{fig:med_energy}a, and the selection $N_\mathrm{channel} > 10$, which removes events that are harder to reconstruct.  
For all energy bins, we apply a reconstructed zenith angle cut of $65\degree$. The only exception is the lowest energy bin, at $E_\mathrm{median}=13$ TeV,  
where we apply a reconstructed zenith angle cut of $60\degree$, since for larger zenith angles, events fall below the $N_{channel}>10$ threshold.  For the highest energy bin, we apply only a lower bound in the energy selection. The primary cosmic-ray energy distributions corresponding to each of the nine selected regions are shown in Fig.~\ref{fig:med_energy}b, along with the associated median energy. Note that the energy distributions of these sets overlap significantly. However, the energy sets are statistically independent as each event is placed in only one of them.
The overlap between the different distributions results mainly from two reasons. First, as in the case of the angular resolution (see Sec.~\ref{ssec:makingmaps}), there is no attempt to optimize the muon energy resolution with dedicated event selections. The goal is to have the largest number of events possible while retaining a reasonable quality. Secondly, the energy deposited by the muon bundles has wide event-by-event fluctuations, and even with a more accurate muon energy estimation, its relation to the parent cosmic-ray particles would still be affected by large fluctuations.

By comparing the H3a~\citep{Gaisser:2012zz}, Polygonato~\citep{horandel}, and GST~\citep{Gaisser:2013dec} composition models, we determined that the uncertainty in the median energy from different cosmic-ray flux models is less than $4$~\% in $\log_{10}(E/\mathrm{GeV})$.
The predicted atmospheric muon fluxes depend on the assumed primary cosmic-ray spectrum and composition and the hadronic models that describe these interactions. Discrepancies among these models lead to an overall value of 20--40\% in the predicted muon flux within the energy range from 1 TeV to 1 PeV~\citep{PhysRevD.102.063002}.

\subsection{Arrival Direction Sky Maps} \label{ssec:maps}
\begin{figure*}[hbtp]
  \centering
  \includegraphics[width=0.49\textwidth]{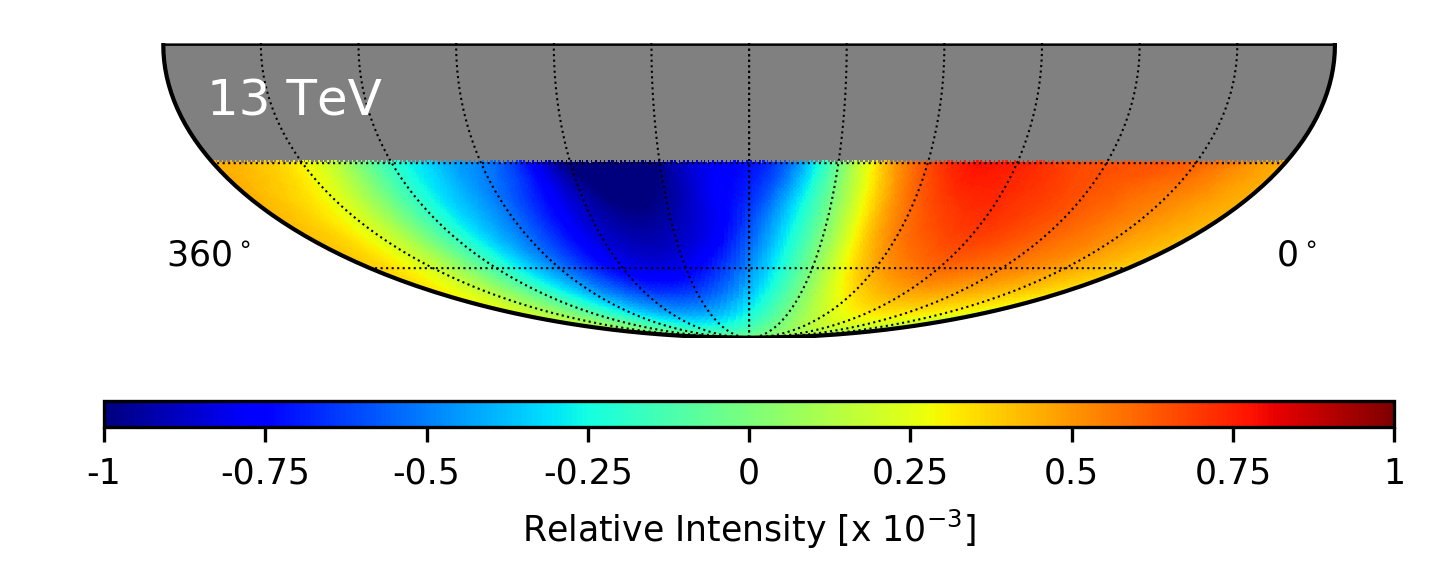}
  \includegraphics[width=0.49\textwidth]{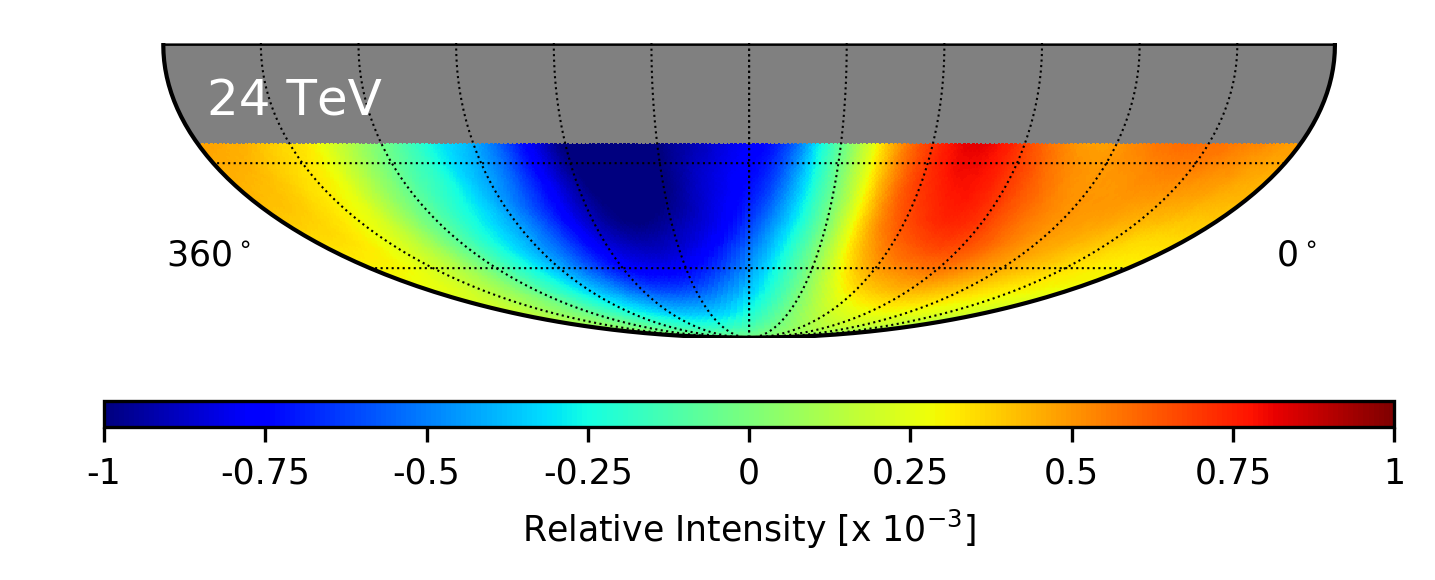}
  \includegraphics[width=0.49\textwidth]{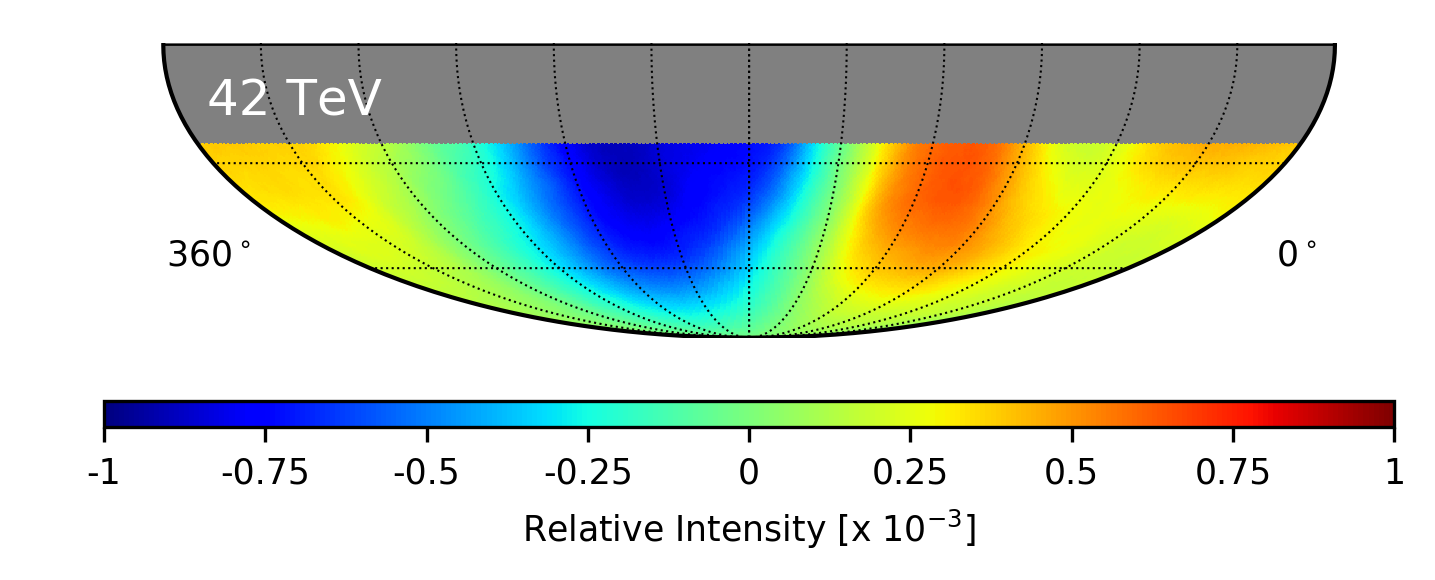}
  \includegraphics[width=0.49\textwidth]{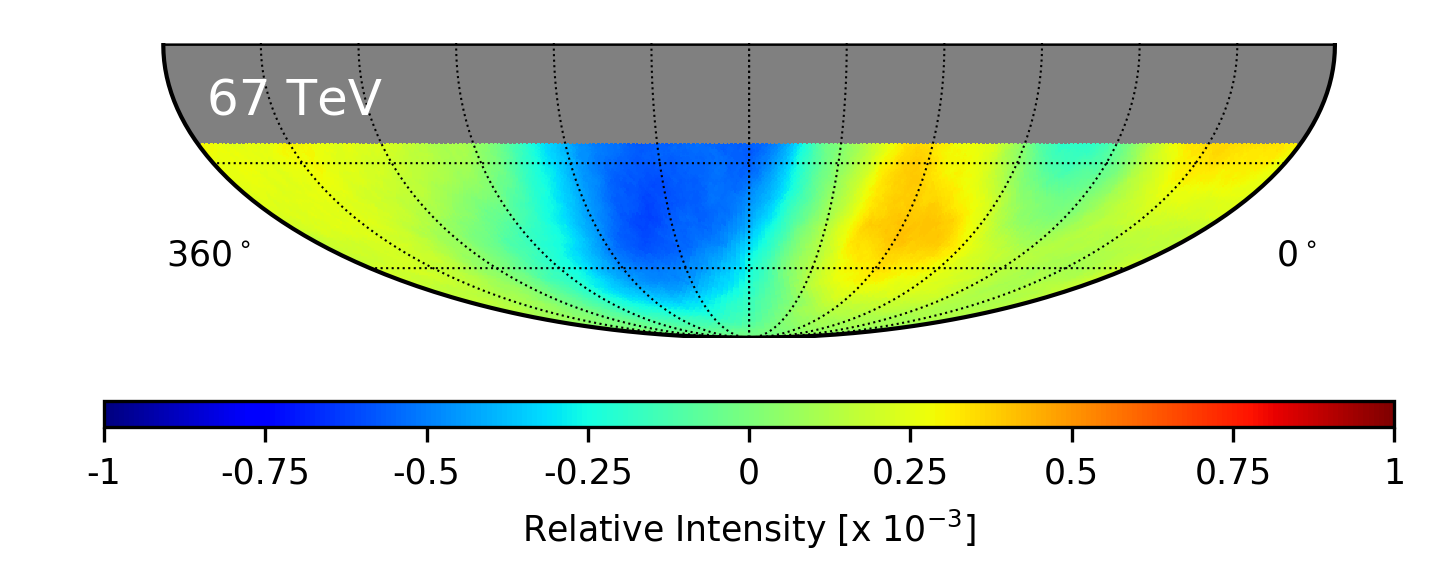}
  \includegraphics[width=0.49\textwidth]{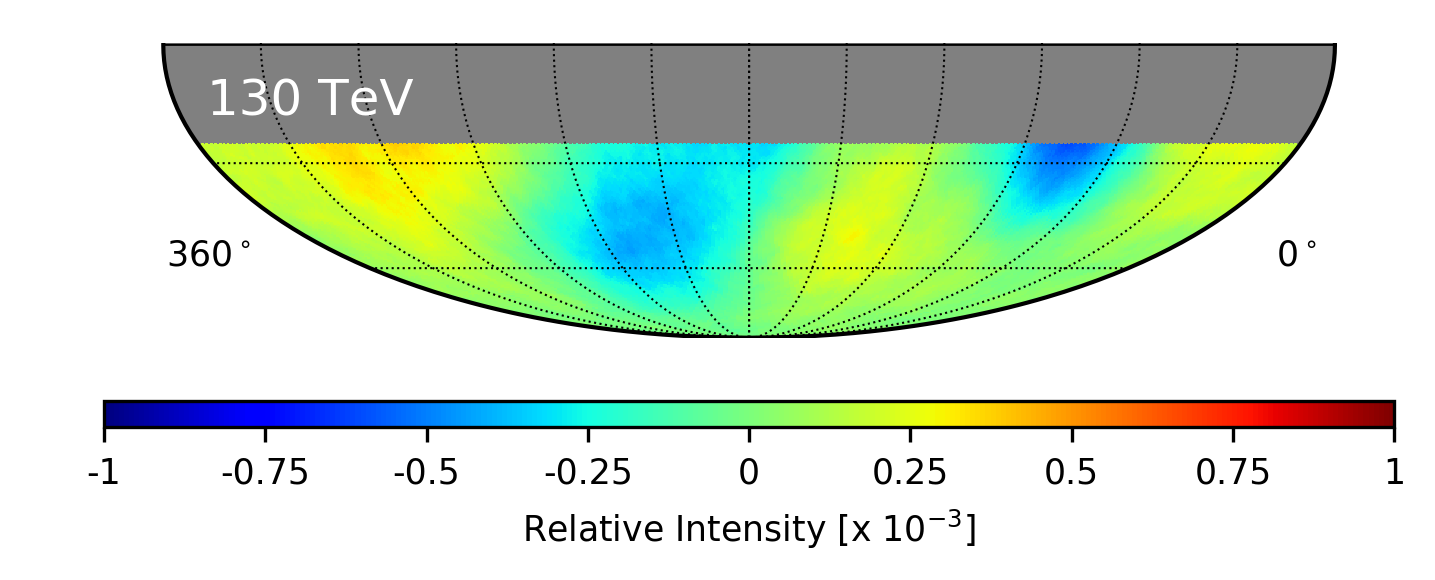}
  \includegraphics[width=0.49\textwidth]{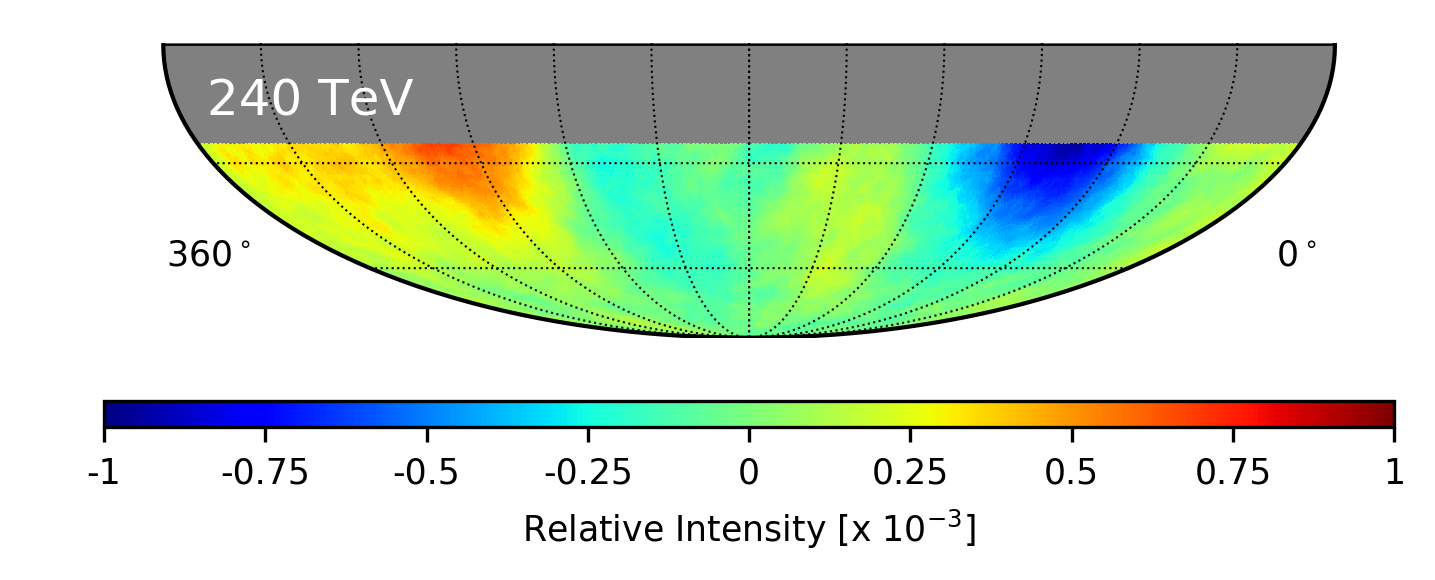}
  \includegraphics[width=0.49\textwidth]{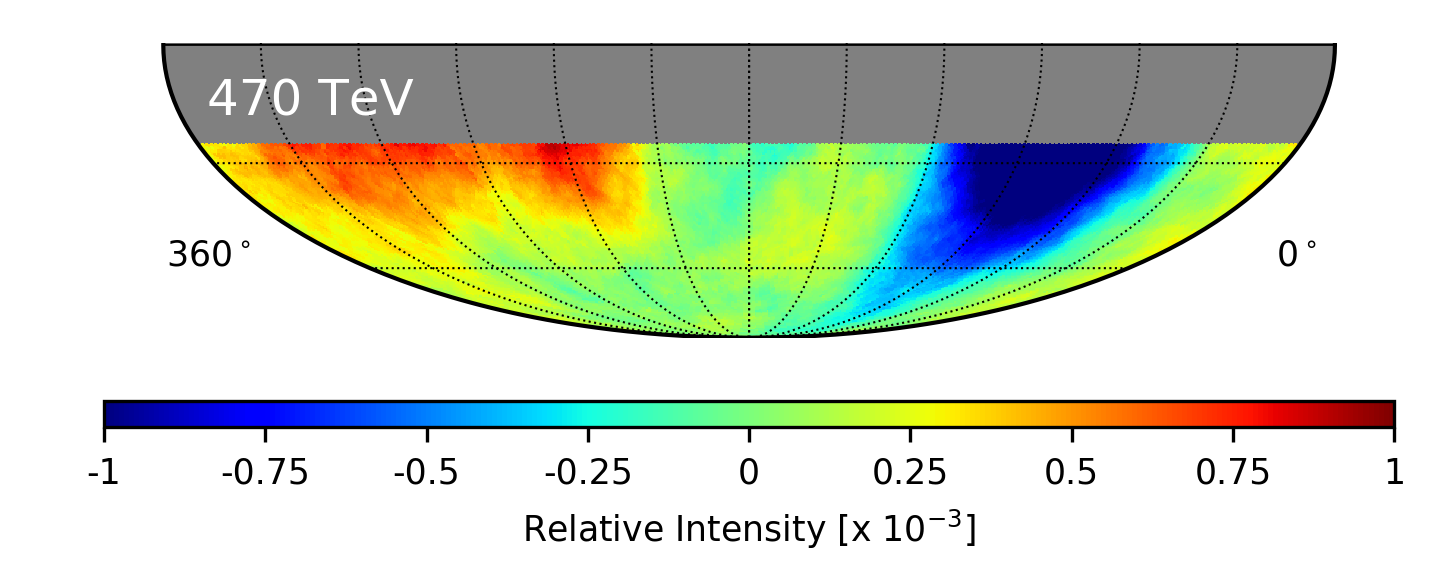}
  \includegraphics[width=0.49\textwidth]{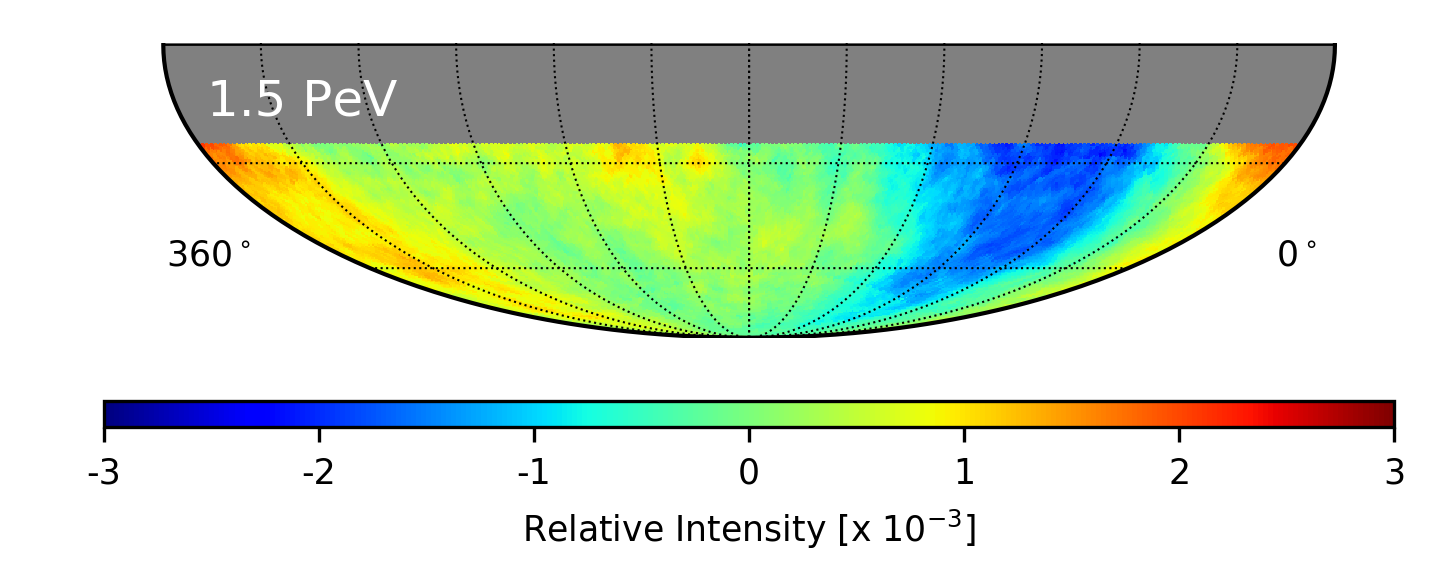}
  \includegraphics[width=0.49\textwidth]{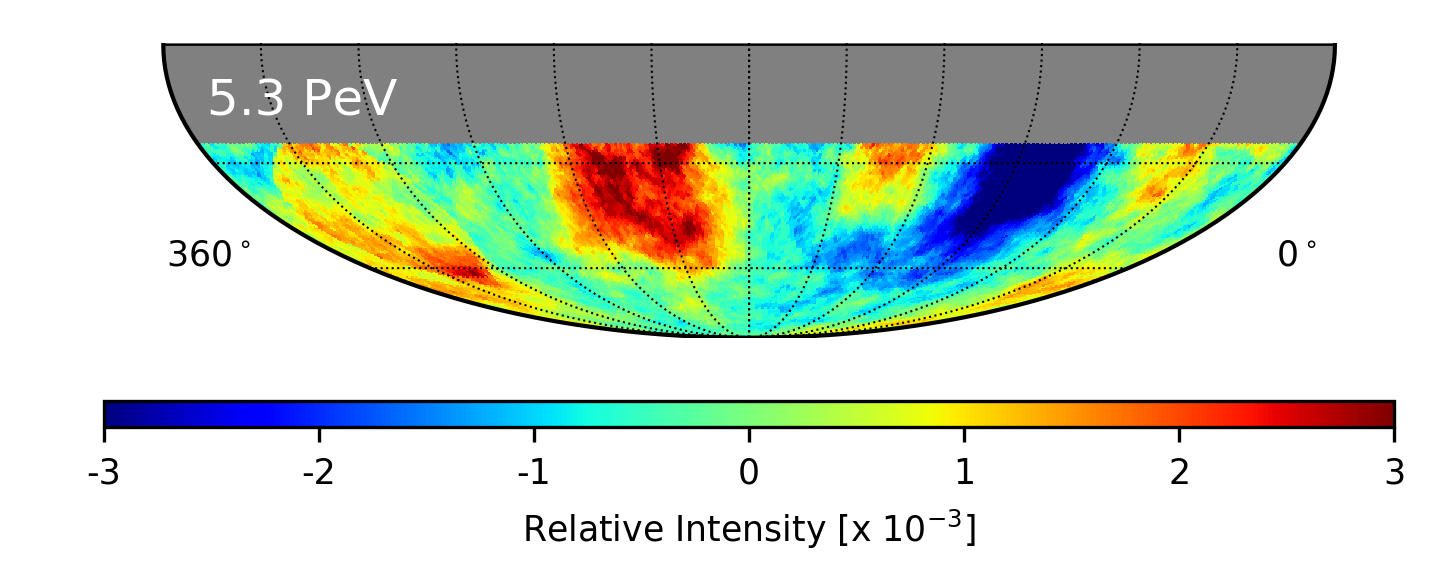}
  \caption{Relative intensity sky maps as a function of primary cosmic-ray energy. The median energy of the data shown in each map is indicated in the upper left. Maps are all in J2000 equatorial coordinates and smoothed with a $20^{\circ}$ smoothing radius for visualization purposes. The final two maps are shown on a different relative intensity scale. The FoV is limited to a zenith angle of 65$^{\circ}$ (-25$^{\circ}$ in equatorial declination). The FoV is further limited to a zenith angle of 60$^{\circ}$ (-30$^{\circ}$ in equatorial declination) for the 13 TeV map.}
  \label{fig:eplots}
\end{figure*}

\begin{figure*}[hbtp]
  \centering
  \includegraphics[width=0.49\textwidth]{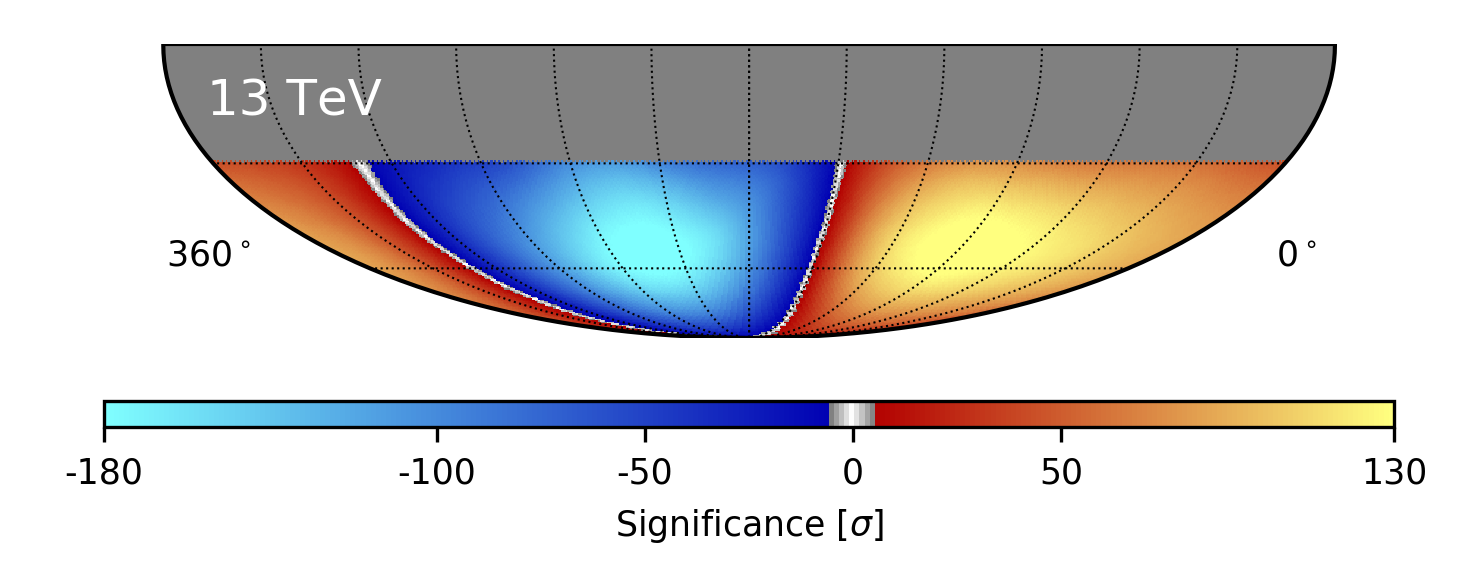}
  \includegraphics[width=0.49\textwidth]{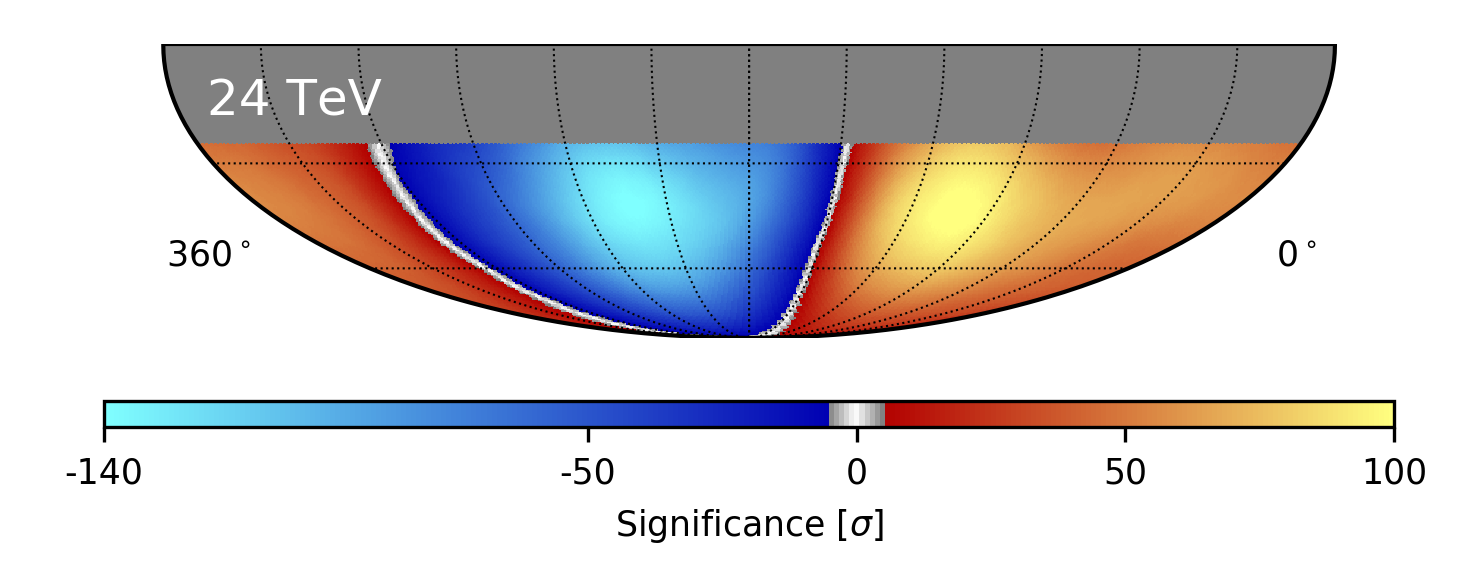}
  \includegraphics[width=0.49\textwidth]{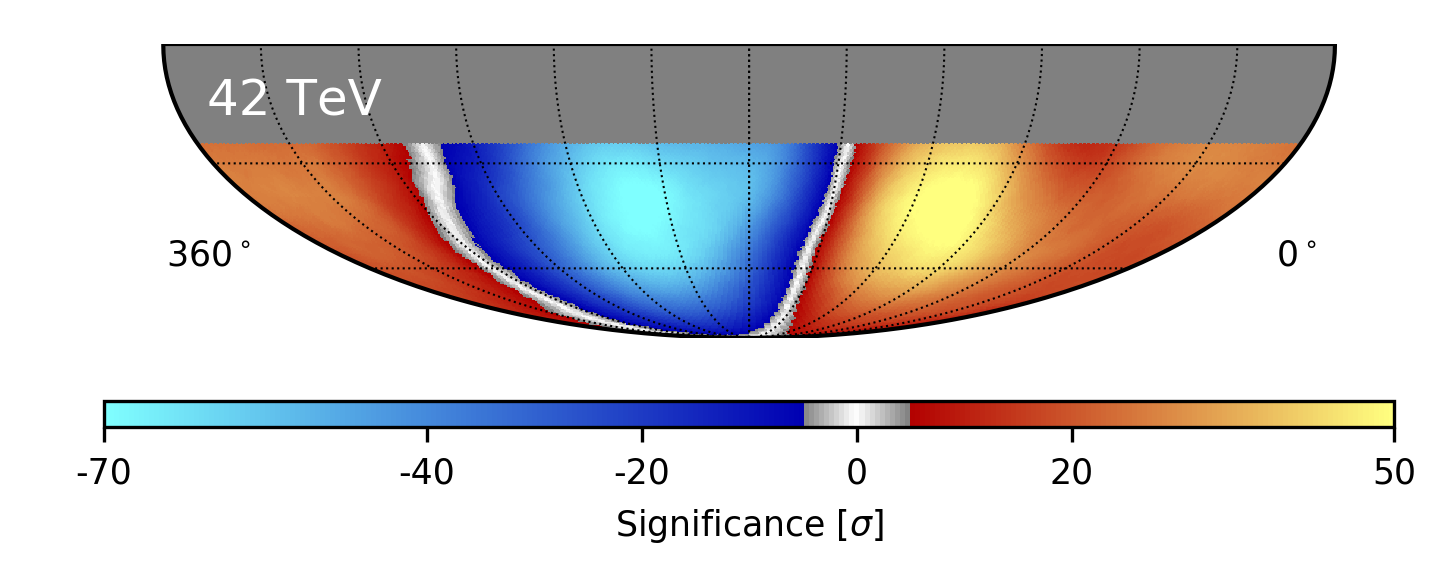}
  \includegraphics[width=0.49\textwidth]{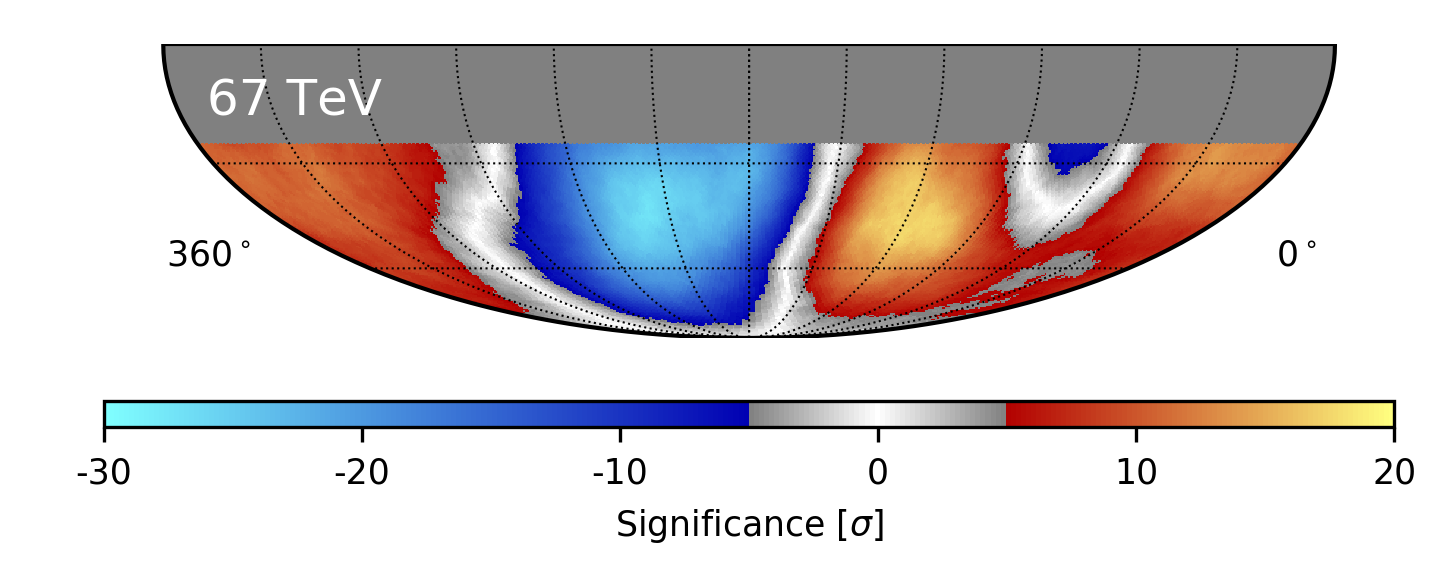}
  \includegraphics[width=0.49\textwidth]{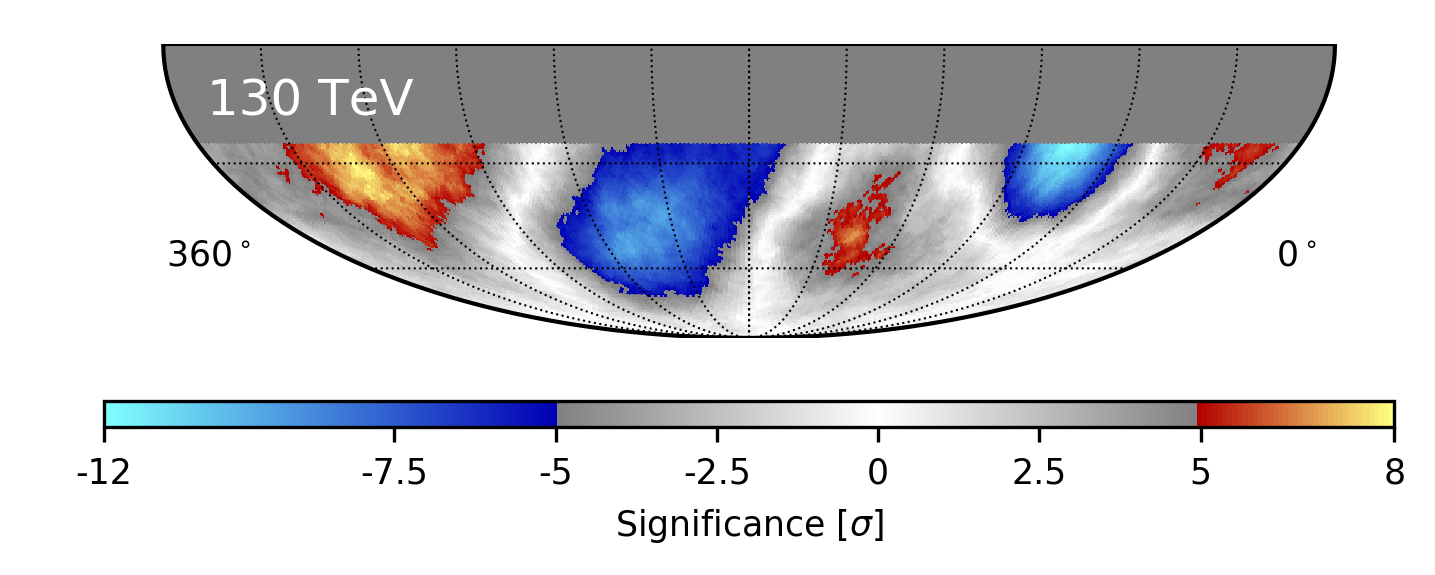}
  \includegraphics[width=0.49\textwidth]{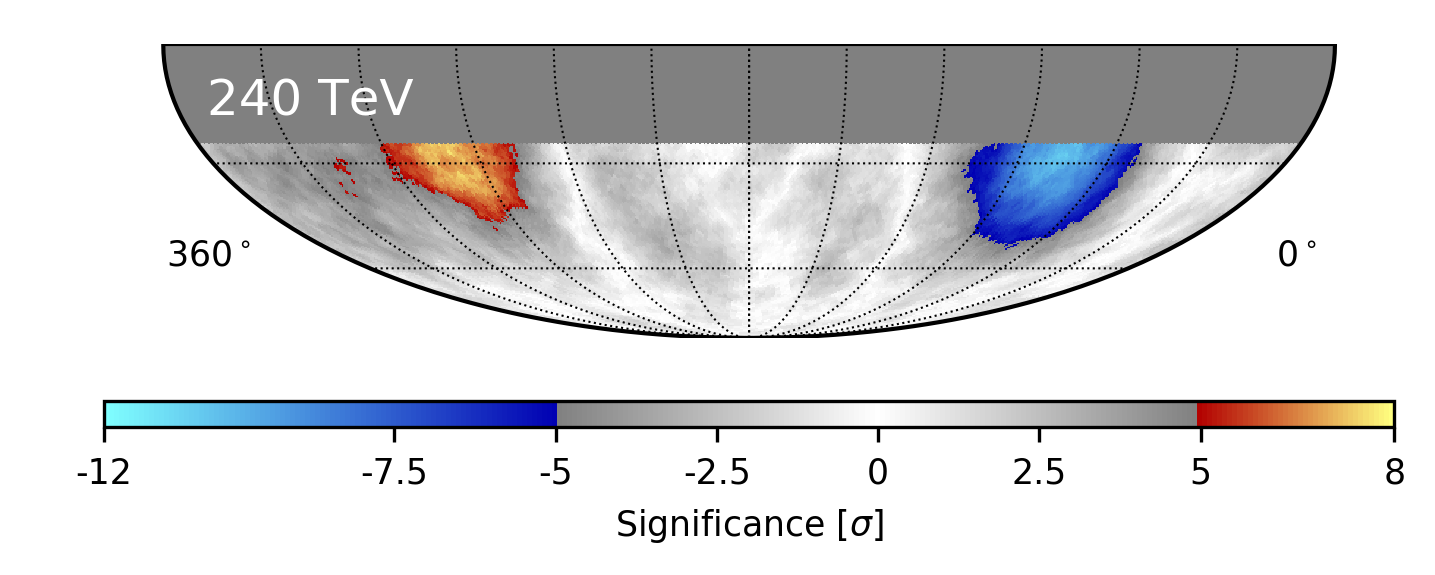}
  \includegraphics[width=0.49\textwidth]{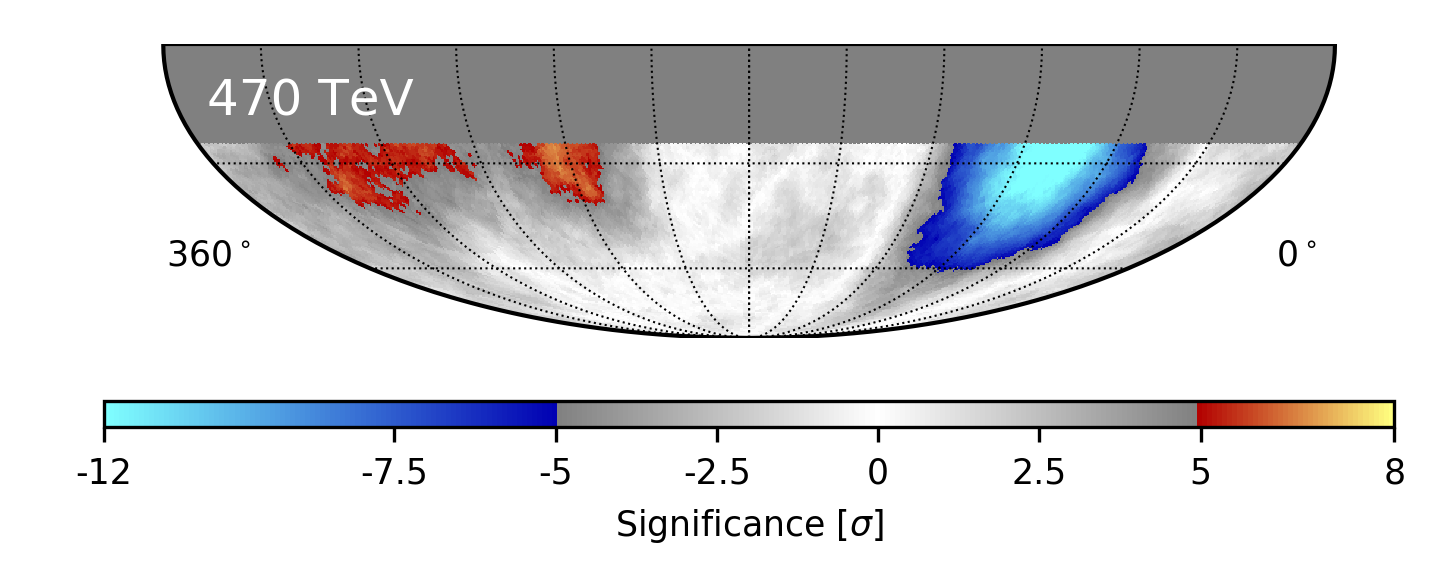}
  \includegraphics[width=0.49\textwidth]{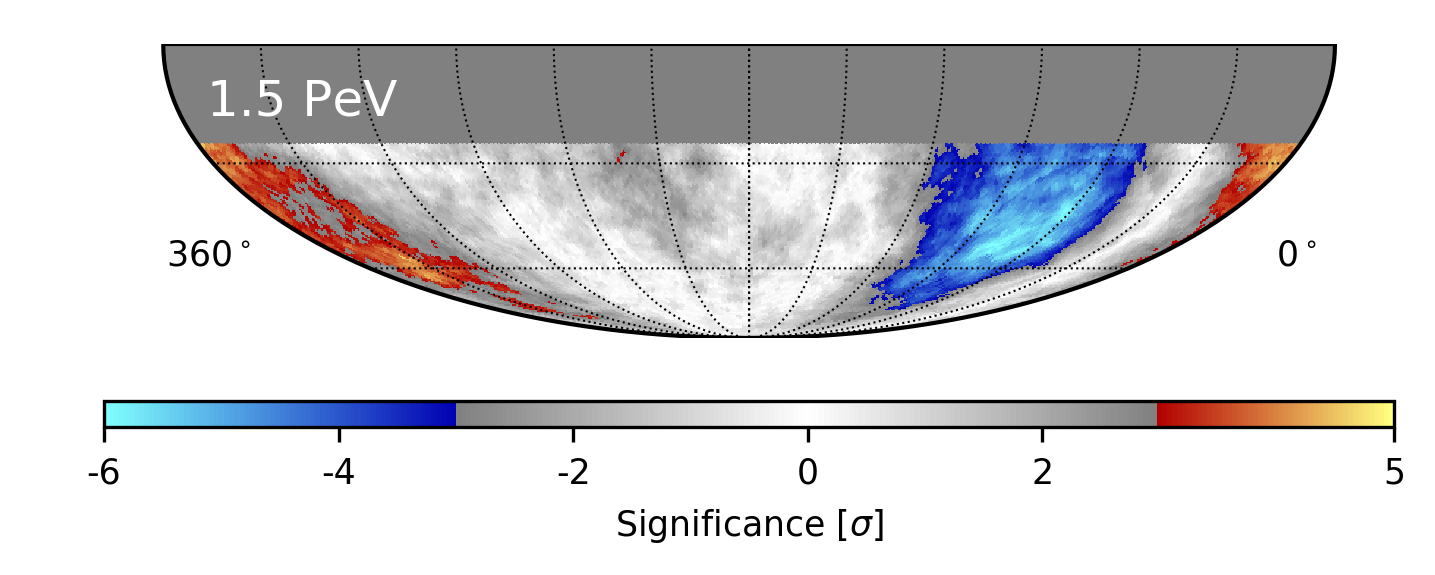}
  \includegraphics[width=0.49\textwidth]{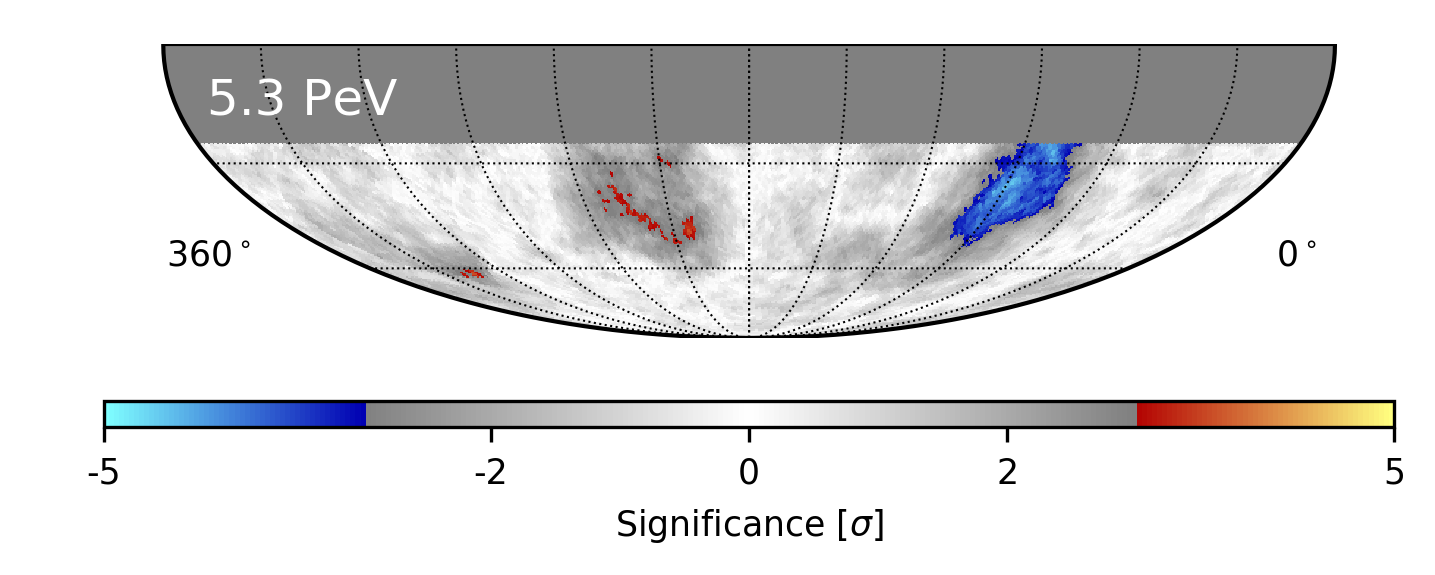}
  \caption{Statistical significance sky maps as a function of primary cosmic-ray energy. The median energy of the data shown in each map is indicated in the upper left. Maps are all in J2000 equatorial coordinates and smoothed with a $20^{\circ}$ smoothing radius for visualization purposes. Features in excess of $5\sigma$ are highlighted ($3\sigma$ in the two highest energy bins). The FoV is limited to a zenith angle of 65$^{\circ}$ (-25$^{\circ}$ in equatorial declination). The FoV is further limited to a zenith angle of 60$^{\circ}$ (-30$^{\circ}$ in equatorial declination) for the 13 TeV map.  }
  \label{fig:eplotssig}
\end{figure*}

We use the sky map calculation procedure described in Sec.~\ref{ssec:makingmaps} on each of the energy data sets described in Sec.~\ref{ssec:makingene} to determine the sky maps of relative intensity and statistical significance shown in Figs.~\ref{fig:eplots} and~\ref{fig:eplotssig}, respectively.

The sky maps show that the observed cosmic-ray anisotropy is on the order of $10^{-3}$, and it dramatically changes in angular structure from 10\,TeV to a few PeV. The directions where the excess/deficit regions are located flip across the sky, with the transition occurring between approximately 100\,TeV and 300\,TeV. It is clearly visible that the cosmic-ray anisotropy cannot be described with a simple dipole (i.e., where the excess and deficit directions are $180^{\circ}$ apart) but rather possesses a complex superposition of different angular scales that evolve with energy.

The statistical significance of the observed cosmic-ray anisotropy sky maps is 
illustrated in Fig.~\ref{fig:eplotssig}. Significance is emphasized in color scale only for regions with $5\sigma$ excess or deficit, except for the two highest energy sky maps (i.e., with median energy of 1.5 and 5.3\,PeV), where it is shown above or below $3\sigma$. The color scheme illustrates deficit significance in blue and excess significance in red. All of these sky maps, except for the one with the highest energy, exhibit large regions of high significance in excess and deficit.

A data release accompanying these results is available in electronic format in~\cite{DVN/DZI2F5_2024}. This release includes the relative intensity sky maps for each energy range used in the analysis (see Fig.~\ref{fig:eplots}), provided in FITS format. The FITS files contain data on event counts, as well as reference event counts, in equatorial coordinates. Additionally, the release features the simulated true particle energy distribution histogram for each of the nine energy sets (refer to Fig.~\ref{fig:med_energy}b).

\vskip 1.cm
\subsection{Characterizing the Dipole Component} \label{ssec:dipole}

The relative intensity $\delta I_i$ in Eq.~\ref{eq:relint} can be expanded in terms of the spherical harmonic functions $Y_{\ell m}$ on the celestial sphere,
\begin{equation}
\label{eq:sphHarmonics}
  {\delta I(\mathbf{\mathbf{u}}_i)} =
    \sum_{\ell=1}^{\infty}
    \sum_{m=-\ell}^{\ell} a_{\ell m} Y_{\ell m}(\mathbf{u}_i) \, ,
\end{equation}
where $\mathbf{\mathbf{u}}_i = (\alpha_i, \delta_i)$; $\alpha_i$ and $\delta_i$ the RA and declination coordinates of the center of pixel $i$, respectively; and $a_{\ell m}$ the multipole coefficients of the expansion.
In the case of partial sky coverage, the standard $Y_{\ell m}$ spherical harmonic functions do not form an orthonormal basis and are highly correlated. An exception is the $m = \ell$ terms, which describe angular structures parallel to the equatorial plane and remain orthogonal to each other under the limited FoV.

\begin{figure*}[t!]
    \centering
    \includegraphics[width=.75\textwidth]{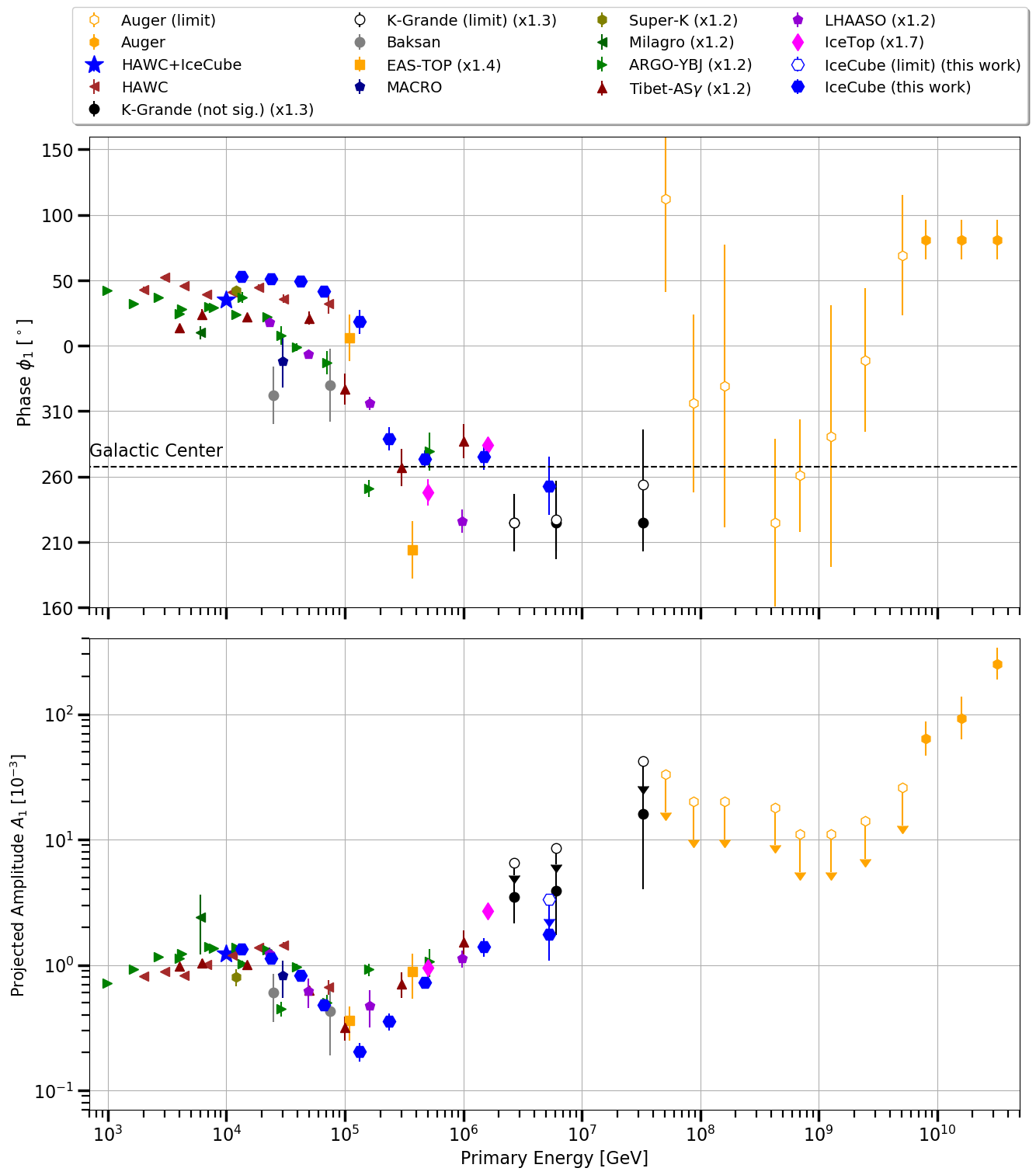}
    \caption{Reconstructed phase (\textit{top}) and amplitude (\textit{bottom}) of the horizontal dipole component for 12 years of IC86 shown along with published measurements from other experiments in the TeV-PeV energy range. 
    The empty symbols represent the 99\% CL upper limits. The RA coordinate of the Galactic Center is indicated as a reference.
    The results shown are from~\cite{
    Abeysekara_2018, 
    Apel_2019, ALEKSEENKO2009179, Aglietta_2009, PhysRevD.67.042002, guillian_2007, Abdo_2009, Bartoli_2015, Bartoli_2018, Amenomori_2005, IceCube:2013mar, Abeysekara_2019, Gao:20211j, 2020ApJ...891..142A}.
Amplitude values for one-dimensional fits have been scaled by factors of 
 1.34 (K-Grande), 1.38 (EAS-TOP),
 1.21 (Super-K), 1.21 (Milagro)
 1.18 (ARGO-YBJ), 1.18 (Tibet-AS$\gamma$)
 1.18 (LHAASO), and 1.74 (IceTop) to account for FoV bias according to the method from \cite{Ahlers:2016rox}. We do not have information about the FoV to determine the scaling factor for the MACRO and Baksan observations. The two-dimensional (HAWC, IceCube) and three-dimensional (Auger) amplitudes do not require correction.}
    \label{fig:dipole}
\end{figure*}

It is common practice to use the dipole component of the anisotropy to compare observations from various cosmic-ray experiments at different energies. To accomplish this, we utilize a simplified version of Eq.~(\ref{eq:sphHarmonics}). We use the two-dimensional function
\begin{equation}
   F(\alpha_i,\delta_i) = \sum\limits_{n=1}^{3} A_n  \cos^n{(\delta_i)}\cos(n \alpha_i + \phi_n) \,,
   \label{eq:2dfit}
\end{equation}
to fit the relative intensity sky maps of Fig.~\ref{fig:eplots}. The $n$ values correspond to the multipole moments $n=\ell=m$, with phase $\phi_n$, and amplitude $A_n$ being proportional to $a_{\ell m}$, in the RA ($\alpha_i$) and declination ($\delta_i$) coordinates associated with pixel $i$. Here we restrict the fit to values with $n = 1,2,3$ (i.e., dipole, quadrupole, and octupole). 
The rationale here is twofold: 
1) The method we use to obtain the reference sky map (see Sec.~\ref{ssec:makingmaps}) effectively averages the event counts over RA to provide the experimental response function vs. declination of the in-ice IceCube array. Therefore, the relative intensity sky maps are not sensitive to cosmic-ray arrival direction modulations across declination bands; events recorded from a fixed direction in the local coordinate system can only probe the cosmic-ray flux at a fixed declination. As a result, the coefficients $a_{\ell 0}$ in Eq.~(\ref{eq:sphHarmonics}) are equal to zero. This can be easily interpreted for the dipole component, where only its projection onto the equatorial plane is measured.
2) In addition, we assume that large-scale features are primarily dominated by $m = \ell$ modes (i.e., parallel to the equatorial plane) extending across the North and South Hemispheres. 
This assumption is somewhat motivated by noting that previous individual measurements in both hemispheres (none sensitive to the $m=0$ terms) observe a horizontal dipole with a similar phase and amplitude, as shown in Fig.~\ref{fig:dipole}.
We find that the IceCube dipole amplitude and phase measurements are approximately in line with the measurements in the north. This implies that, particularly in the case of the quadrupole, the $a_{2 m}$ terms with $m=\pm 1$ are negligible; otherwise, the dipole amplitudes and phases in the north and south would differ significantly.
Table~\ref{tab:1damp} shows the amplitude $A_1$ and phase $\phi_1$ of the dipole component from the fit for each of the nine energy bins with the two-dimensional function Eq.~\ref{eq:2dfit}. 
\begin{table*}[t!]
    \footnotesize
    \centering
\begin{tabular}{l || rcl | rcl | rcl || rcl | rcl}
  $E_\mathrm{med}$ [$\mathrm{TeV}$] & \multicolumn{3}{c}{$A_1[10^{-4}]$}
                                    & \multicolumn{3}{c}{$\tilde{A}_1 [10^{-4}]$} 
                                    & \multicolumn{3}{c}{$A^\dagger_{1} [10^{-4}]$} 
                                    & \multicolumn{3}{c}{$\phi_1 [^\circ]$} 
                                    & \multicolumn{3}{c}{$\tilde{\phi}_1 [^\circ]$} \\
\hline
\hline
13   & 13.33 &$\pm$& 0.05  &  7.552 & $\pm$ & 0.028 & 13.14 & $\pm$ & 0.05 &  52.76 & $\pm$ &  0.20 &  52.40 & $\pm$ &  0.21 \\
24   & 11.24 &$\pm$& 0.05  &  7.096 & $\pm$ & 0.035 & 11.55 & $\pm$ & 0.06 &  50.91 & $\pm$ &  0.28 &  51.30 & $\pm$ &  0.28 \\
42   &  8.28 &$\pm$& 0.09  &  5.13  & $\pm$ & 0.05  &  8.36 & $\pm$ & 0.09 &  49.2  & $\pm$ &  0.6  &  49.3  & $\pm$ &  0.6  \\
67   &  4.81 &$\pm$& 0.17  &  3.27  & $\pm$ & 0.11  &  5.33 & $\pm$ & 0.19 &  41.5  & $\pm$ &  2.0  &  44.8  & $\pm$ &  2.0  \\
130  & 2.04  &$\pm$& 0.33  &  1.26  & $\pm$ & 0.22  &  2.1  & $\pm$ & 0.4  &  18    & $\pm$ &  9    &  28    & $\pm$ & 10    \\
240  & 3.6   &$\pm$& 0.5   &  2.0   & $\pm$ & 0.4   &  3.3  & $\pm$ & 0.6  & 289    & $\pm$ &  9    & 291    & $\pm$ & 10    \\
470  & 7.2   &$\pm$& 0.8   &  4.3   & $\pm$ & 0.5   &  7.0  & $\pm$ & 0.8  & 274    & $\pm$ &  6    & 277    & $\pm$ &  7    \\
1500 & 14.0  &$\pm$& 2.4   &  9.2   & $\pm$ & 1.6   & 15.0  & $\pm$ & 2.5  & 275    & $\pm$ & 10    & 278    & $\pm$ & 10    \\
5300 & 18    &$\pm$& 7     & 10     & $\pm$ & 4     & 16    & $\pm$ & 7    & 253    & $\pm$ & 22    & 263    & $\pm$ & 25    \\
\end{tabular}
    \vspace{0.1in}
    \caption{\small Amplitudes and phases of the dipole component for each of the nine energy sets. The values $A_1$ and $\phi_1$ are from the two-dimensional fit (Eq.~\ref{eq:2dfit}). The values $\tilde{A}_1$ and $\tilde{\phi}_1$ are from the fit to the one-dimensional projection. The $A^\dagger_{1}$ are the amplitudes with the geometric correction to account for the limited FoV bias from \cite{Ahlers:2016rox}. The first column corresponds to the median energy of each energy set.}
    \label{tab:1damp}
\end{table*}

Previous IceCube results of the dipole amplitude 
were determined from the declination average of the relative CR intensity~$I$ (i.e., one-dimensional projection). As noted by \cite{Ahlers:2016rox},  
this results in an underestimation of the actual dipole amplitude, equivalent to Eq.~\ref{eq:2dfit} averaged over declination between the minimum ($\delta_\mathrm{min}$) and maximum ($\delta_\mathrm{max}$) values,
\begin{align}\label{eq:1dfit}
\begin{split}
\tilde{F}(\alpha_i) & = \frac{1}{\sin{\delta_\mathrm{max}}-\sin{\delta_\mathrm{min}}}\int\limits_{\delta_\mathrm{min}}^{\delta_\mathrm{max}}F(\alpha_i,\delta)\cos{\delta}\, d\delta \\
& = \sum\limits_{n=1}^{3} \tilde{A}_n  \cos(n \alpha_i + \tilde{\phi}_n)\; ,
\end{split}
\end{align}
where $\tilde{A}_n$ and $\tilde{\phi}_n$ are the amplitudes and phases of the one-dimensional measurement.

For the dipole component, Eq.~\ref{eq:1dfit} reduces to
\begin{widetext}
\begin{align}  
\label{eq:corr1dfit}
\begin{split}
\tilde{F}(\alpha_i) &= \frac{A_1}{\sin{\delta_\mathrm{max}}-\sin{\delta_\mathrm{min}}}\int\limits_{\delta_1}^{\delta_2}\cos^2{\delta}\cos{(\alpha_i+\phi_1)}d\delta \\
& = \frac{\delta_\mathrm{min}-\delta_\mathrm{max} + \cos\delta_\mathrm{min}\sin\delta_\mathrm{min} - \cos\delta_\mathrm{max}\sin\delta_\mathrm{max}}{2(\sin\delta_\mathrm{min}-\sin\delta_\mathrm{max})}A_1 \cos{(\alpha_i+\phi_1)} \\
& = \tilde{A}_1 \cos{(\alpha_i + \tilde{\phi}_1)}\; .
\end{split}
\end{align}
\end{widetext}
From Eq. \ref{eq:corr1dfit}, we obtain a geometric scaling factor, 
\begin{widetext}
\begin{equation}\label{eq:1dcorr}
    A^\dagger_{1} \equiv \frac{2(\sin\delta_\mathrm{min}-\sin\delta_\mathrm{max})}{\delta_\mathrm{min}-\delta_\mathrm{max} + \cos\delta_\mathrm{min}\sin\delta_\mathrm{min} - \cos\delta_\mathrm{max}\sin\delta_\mathrm{max}}\tilde{A}_1 \; ,
\end{equation}
\end{widetext}
where $A^\dagger_{1}$ is the one-dimensional measurement of the dipole amplitude, $\tilde{A}_1$, rescaled to account for the limited FoV of the dipole projected on the equatorial plane, and is equivalent to $A_1$, obtained from the two-dimensional fit of Eq.~\ref{eq:2dfit}.
Note that the two-dimensional function is preferred to the one-dimensional function because it naturally accounts for the experiment's limited FoV, not because it is expected to provide a better fit. 
This is shown in Table~\ref{tab:1damp}, where the scaled $A^\dagger_{1}$ values obtained from Eq.~\ref{eq:1dcorr} agree with the two-dimensional fit within statistical uncertainties. We exclude pixels near the horizon where the angular resolution worsens to improve the fit. We select only pixels with $\delta < \delta_\mathrm{max}$, with $\delta_\mathrm{max} =-35^\circ$ for the 13\,TeV map, due to the larger $\cos{\theta}$ threshold (see Fig. \ref{fig:med_energy}), and $\delta_\mathrm{max}=-30^\circ$ for all other energy bins. 

Figure~\ref{fig:dipole} shows IceCube's dipole anisotropy components as a function of energy (obtained from the fit of Eq.~\ref{eq:2dfit} to the data) alongside the results from other experiments. The data points for experiments that use a one-dimensional projection are scaled by the geometric factor from Eq.~\ref{eq:1dcorr} obtained from their respective FoVs. 
The figure quantitatively shows what is visually evident in the sky maps of Figs.~\ref{fig:eplots} and~\ref{fig:eplotssig}. The dipole component's amplitude decreases from 10\,TeV to slightly above 100\,TeV, reaching a minimum of about $2\times 10^{-4}$ where the phase flips from an RA of about $50^{\circ}$ to about $260^{\circ}$. The amplitude increases again at higher energy, generally similar to the observations of all other ground-based experiments. 

Because the dipole fit to the IceCube data at 5.3\,PeV is only marginally distinguishable (i.e., at the level of 1.5$\sigma$) from the isotropy hypothesis (see also Fig.~\ref{fig:eplotssig}), we calculated the corresponding upper limit. To do this, we used the data count sky map of the 5.3\,PeV event sample, generated 100,000 maps, with each pixel containing a Poisson-randomized count with a mean value taken from the data map, and computed the two-dimensional fit of the relative intensity with Eq.~\ref{eq:2dfit} for each of the samples. The 100,000 dipole amplitude values were then used to construct a probability density distribution. The upper limit on the amplitude corresponding to the 99\% CL is $A_1^{99}(5.3\,\mathrm{PeV})\leq 3.34\times 10^{-3}$. In Fig.~\ref{fig:dipole}, upper limits on the dipole amplitude after the scaling factor to account for the FoV are reported with empty symbols. 

\begin{figure}[t!]
    \centering
    \includegraphics[width=.95\columnwidth]{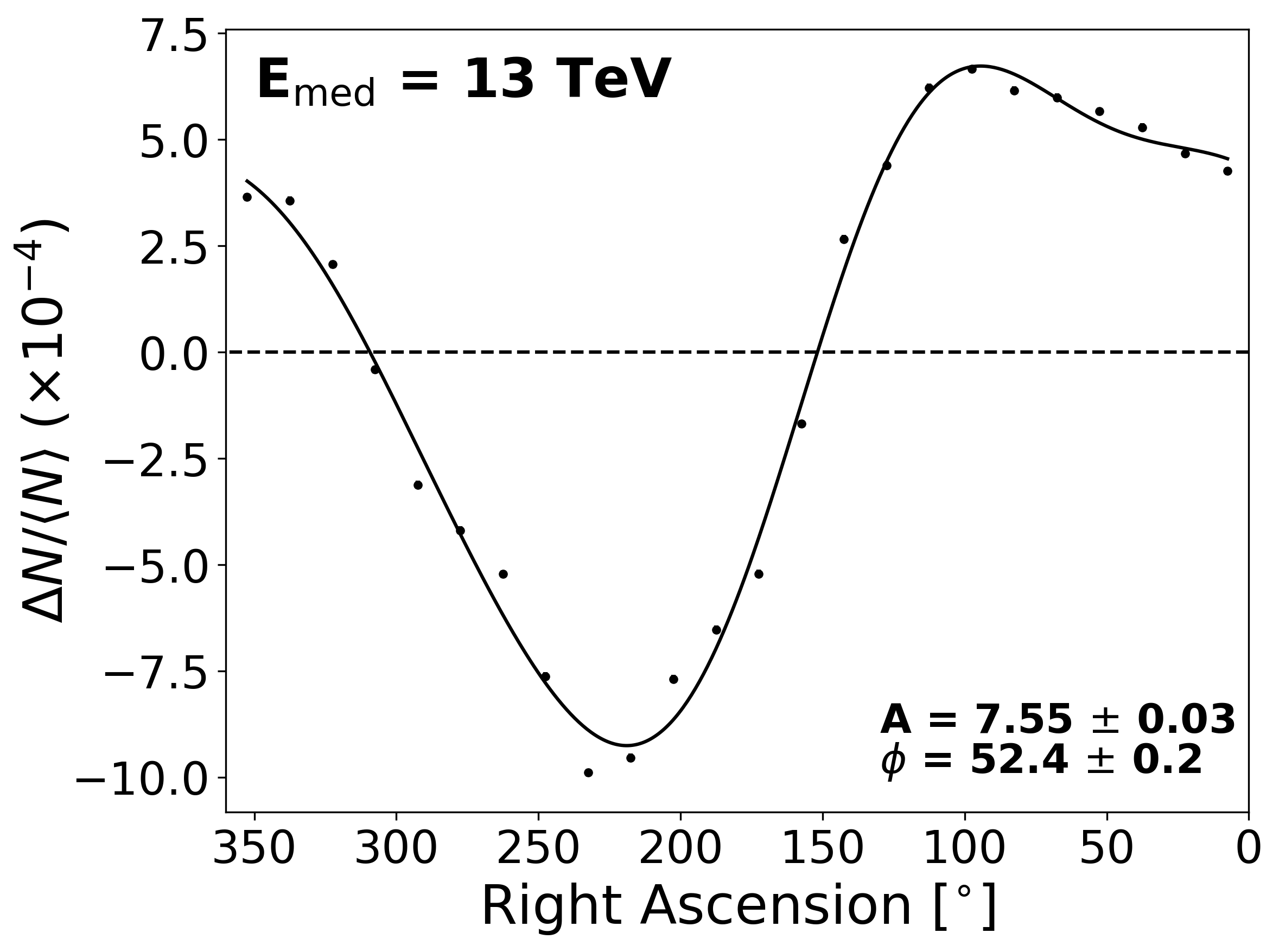}
    \caption{The one-dimensional projection of the relative intensity for the 13 TeV median energy data sample (points), and the corresponding fit to the one-dimensional harmonic function eq.~\ref{eq:1dfit} (solid curve). Note that the statistical errors are approximately the markers' size.}
    \label{fig:1dfitplot}
\end{figure}
The quality of the fits to the harmonic functions (Eqs. \ref{eq:2dfit}, \ref{eq:1dfit}) described in this section is relatively poor because higher multipole terms would be required to fully characterize the observed distributions, as illustrated in Fig.~\ref{fig:1dfitplot}. However, the first three harmonics terms are sufficient to describe the overall relative intensity well in the one-dimensional case. In fact, we find that the dipole component measurement is stable well within the fit uncertainty, independently of how many multipole terms we use in the fit. This is consistent with the fact that the $m = \ell$ terms are orthogonal, meaning that the dipole term is unaffected by the presence of higher-order multipole terms, and it is a stable and reliable measurement even when the overall fit quality is not optimal.

\subsection{Angular Power Spectrum} \label{ssec:ang}

To quantify how different angular scales observed in the relative intensity sky maps contribute to the overall anisotropy, we calculate the angular power spectrum for each of the maps in Fig.~\ref{fig:eplots} using Eq.~\ref{eq:sphHarmonics} via healpy's \texttt{anafast} method~\citep{Zonca2019, Gorski_2005,Zonca2019}. 
Unlike the determination of the anisotropy's dipole component in Sec.~\ref{ssec:dipole}, here we perform a full harmonic decomposition to calculate the angular power spectrum.
In the case of full-sky coverage over $4\pi$ steradians, the multipole moments ${a}_{\ell m}$ fully describe the relative intensity distribution. 
\begin{figure*}[hbtp]
  \centering
  \includegraphics[width=0.49\textwidth]{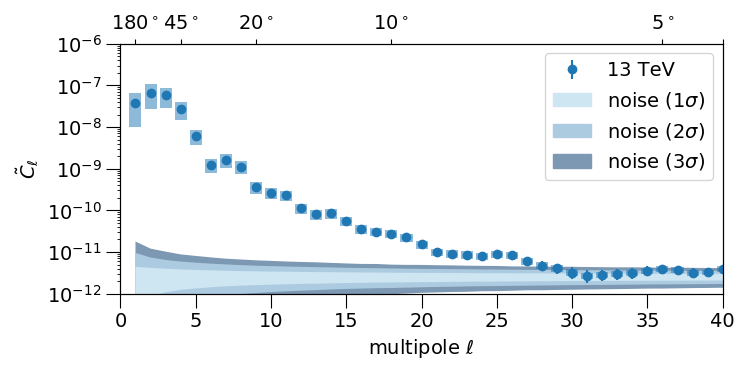}
  \includegraphics[width=0.49\textwidth]{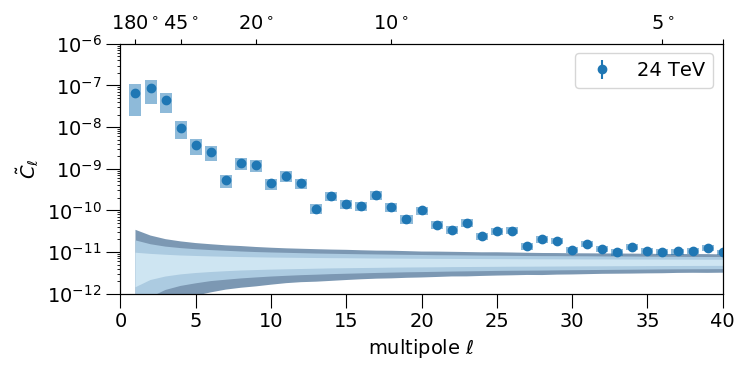}
  \includegraphics[width=0.49\textwidth]{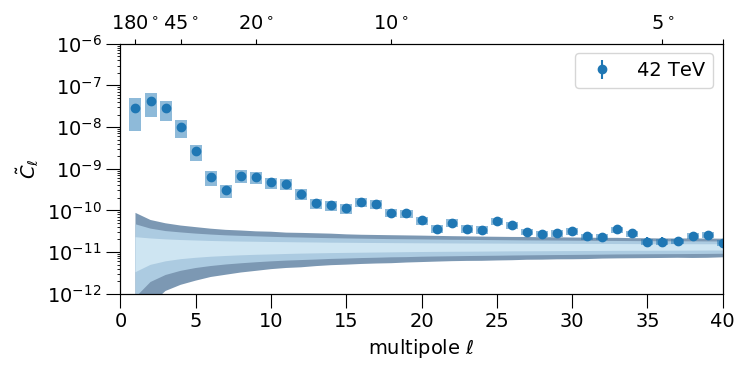}
  \includegraphics[width=0.49\textwidth]{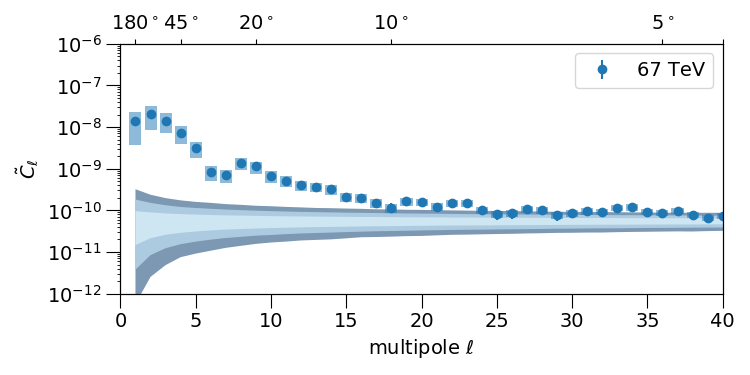}
  \includegraphics[width=0.49\textwidth]{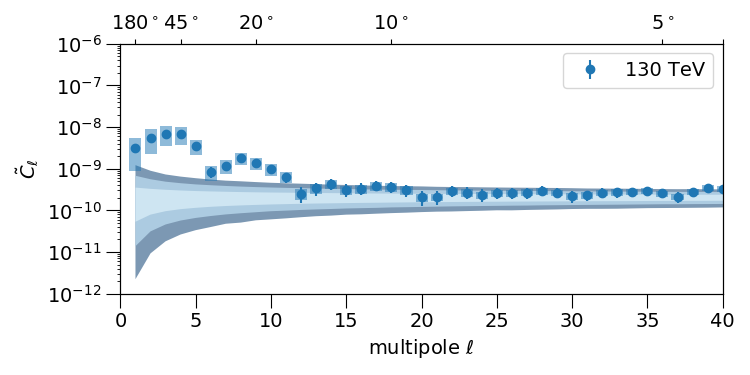}
  \includegraphics[width=0.49\textwidth]{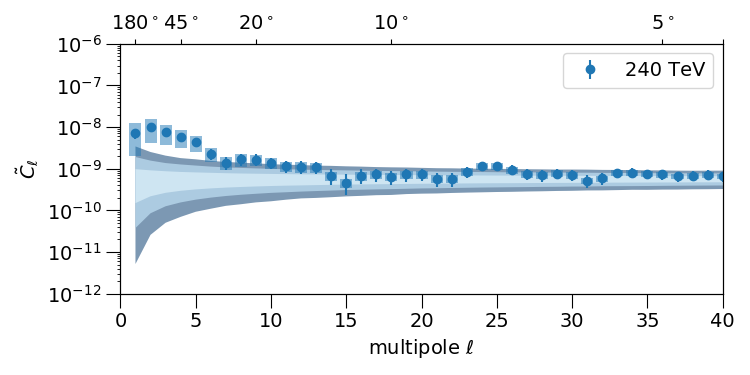}
  \includegraphics[width=0.49\textwidth]{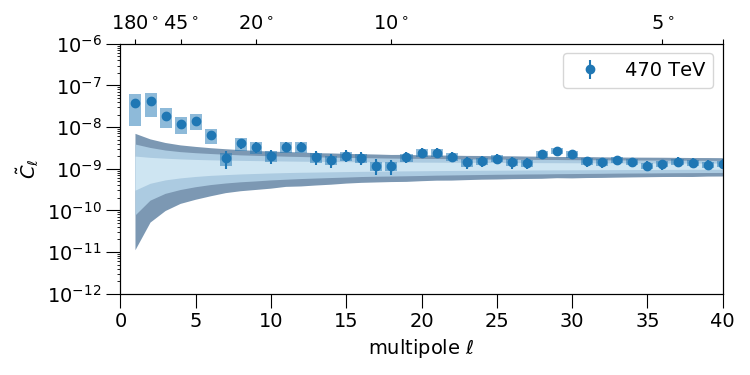}
  \includegraphics[width=0.49\textwidth]{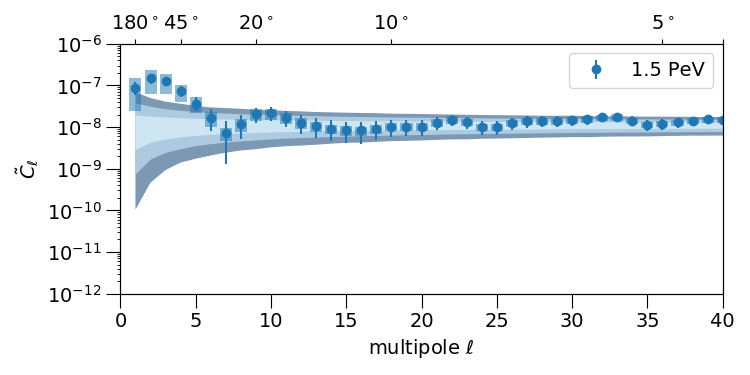}
  \caption{Angular pseudo-power spectra as a function of energy. The shaded boxes represent systematic uncertainties from calculating the power spectrum; the other error bars are statistical, produced by calculating power spectra for Poisson-fluctuated data maps. The large, blue bands indicate the 68\%, 95\%, and 99.7\% spread in $\tilde{C}_{\ell}$ for a large sample of scrambled maps. The highest energy bin is omitted, as its power spectrum is consistent with an isotropic background at all multipoles.}
  \label{fig:ebinned_aps}
\end{figure*}
The measured angular power spectrum of the relative intensity map corresponds to the variance of the $a_{\ell m}$ coefficients 
\begin{equation}
  \mathcal{C}_{\ell} = \frac{1}{2 \ell + 1} \sum_{m=-\ell}^{\ell} | a_{\ell m} |^{2}
   \label{eq:cldef_true}
\end{equation}
of the $Y_{\ell m}(\mathbf{\mathbf{u}})$ terms in the spherical harmonic expansion (Eq. \ref{eq:sphHarmonics})
for each value of $\ell$~\citep{Scott_2016}, and provides an estimate of the strength of structures at different angular scales of $\sim$ 180$^\circ/\ell$.

However, the restricted sky coverage of this analysis results in multipole  coefficients that are highly correlated
between different $\ell$ modes. This correlation is greater for small values of $\ell$ but decreases with increasing $\ell$~\citep{SOMMERS2001271, Abeysekara_2019}.
We henceforth refer to the biased fit coefficients as $\hat{a}_{\ell m}$ to distinguish them from the true, unbiased $a_{\ell m}$. 

As mentioned in Sec.~\ref{ssec:dipole}, this analysis is not sensitive to the component of the arrival directions distribution perpendicular to the equatorial plane (i.e., the north-south direction). 
In fact, we find that the values of the corresponding reconstructed coefficients $\hat{a}_{\ell 0}$ are approximately $10^{-10}$, consistent with the fact that our measurement is not sensitive to the north-south anisotropy component
(see Sec.~\ref{ssec:sysaps} for additional information). Therefore, we calculate what we refer to as an angular \textit{pseudo-power} spectrum $\mathcal{\tilde{C}}_{\ell}$ defined as
\begin{equation}
  \mathcal{\tilde{C}}_{\ell} = \frac{1}{2 \ell + 1} \sum_{m=-\ell, m\neq 0}^{\ell} | \hat{a}_{\ell m} |^{2}~~.
   \label{eq:cldef}
\end{equation}

Figure~\ref{fig:ebinned_aps} shows the angular pseudo-power spectra for each of the energy sets described in Sec.~\ref{ssec:makingene} (except the highest energy). 
A study of how a limited FoV affects the angular power spectrum shape is discussed in Sec.~\ref{ssec:sysaps}.

The angular pseudo-power spectra in Fig.~\ref{fig:ebinned_aps} are compared to the corresponding reference maps that describe IceCube's response to an isotropic cosmic-ray flux (see Sec.~\ref{ssec:maps}). We calculated the angular power spectra for 100,000 Poisson-fluctuated reference maps obtained with time-scrambling for each energy bin. The 1$\sigma$, 2$\sigma$, and 3$\sigma$ spread distributions of the angular power spectrum at each value of $\ell$ are shown as shaded bands in Fig.~\ref{fig:ebinned_aps}, and they represent the statistical \textit{noise} originated from Poisson fluctuations of an isotropic sky map. Note that the noise power increases with energy due to the decrease in the number of events. 

\begin{figure*}[h!]
    \centering
    \includegraphics[width=.99\linewidth]{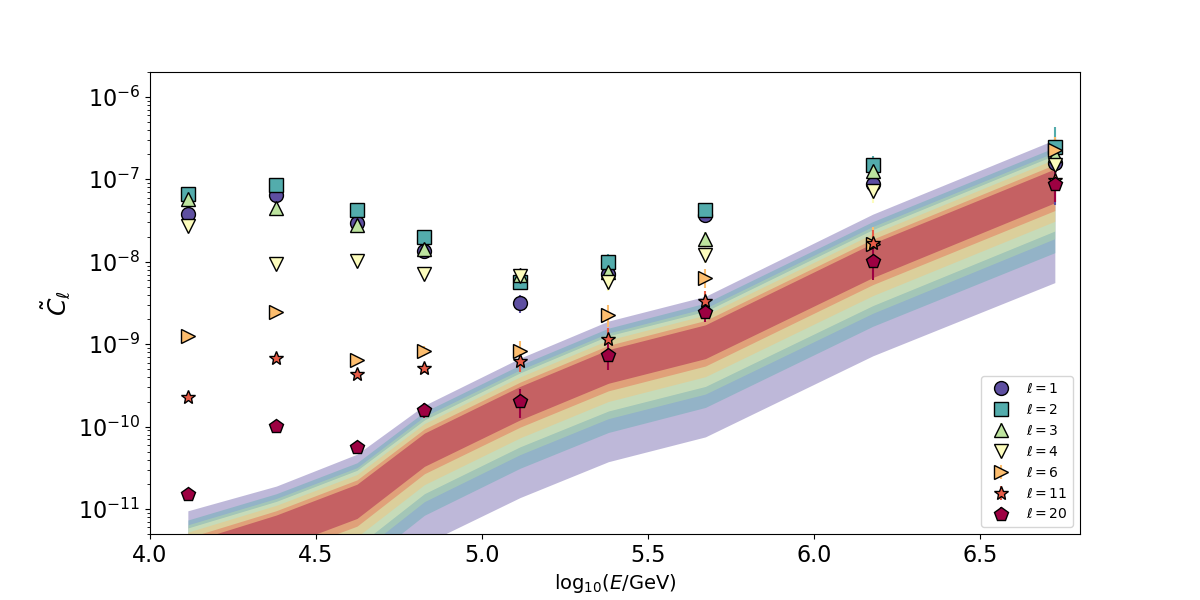}
    \caption{Angular pseudo-power for selected spherical harmonic modes ($\ell$) as a function of median primary energy for twelve years of IceCube data. Errors bars are statistical. 
    The bands indicate the 95\% spread in $\tilde{C}_{\ell}$ from a large sample of scrambled maps. The color of the bands corresponds to that of the data symbols but in a lighter shade.}
    \label{fig:apse}
\end{figure*}

To calculate statistical uncertainties (the solid line bars in Fig.~\ref{fig:ebinned_aps} visible for large values of $\ell$), we looked at the spread of angular power spectra produced by Poisson-fluctuated data maps. Systematic uncertainties in the harmonic decomposition (the shaded boxes in Fig.~\ref{fig:ebinned_aps}) were estimated by producing false sky maps with the same $\tilde{C}_{\ell}$ but random $a_{\ell m}$ from our observed power spectrum (using healpy's \texttt{synfast} method) and subsequently recalculating the power spectra. This round-trip test estimates the stability of the \texttt{anafast} algorithm in reconstructing the angular power spectrum.
In both cases, our error bars represent the 68\% spread around the average for each multipole term. 

Figure~\ref{fig:apse} shows the angular pseudo-power spectra as a function of the median primary cosmic-ray energy for different values of $\ell$. This representation is an extension of Fig.~\ref{fig:ebinned_aps} to visualize how the angular power changes with energy. The dipole pseudo-power behaves similarly to the amplitude in Fig.~\ref{fig:dipole}. Note that the multipole terms $\ell \simeq 1-3$ show a similar energy dependence.

\subsection{Angular Scale Decomposition} \label{ssec:small-scale}

The sky maps in Figs.~\ref{fig:eplots} and \ref{fig:eplotssig} show a complex superposition of different angular scales evolving with energy. At energies below approximately 100 TeV, the anisotropy appears smooth and exhibits more wide-angle features compared to maps at higher energy. This observation is supported by the decline in the pseudo-power $\tilde{C}_\ell$ for multipole terms with $\ell \leq 3$ up to about 100\,TeV, as seen in Fig.~\ref{fig:apse}. Conversely, the structures with $\ell \simeq 6-11$ appear relatively stable as a function of energy.

\begin{figure*}[t!]
  \centering
  \includegraphics[width=0.49\textwidth]{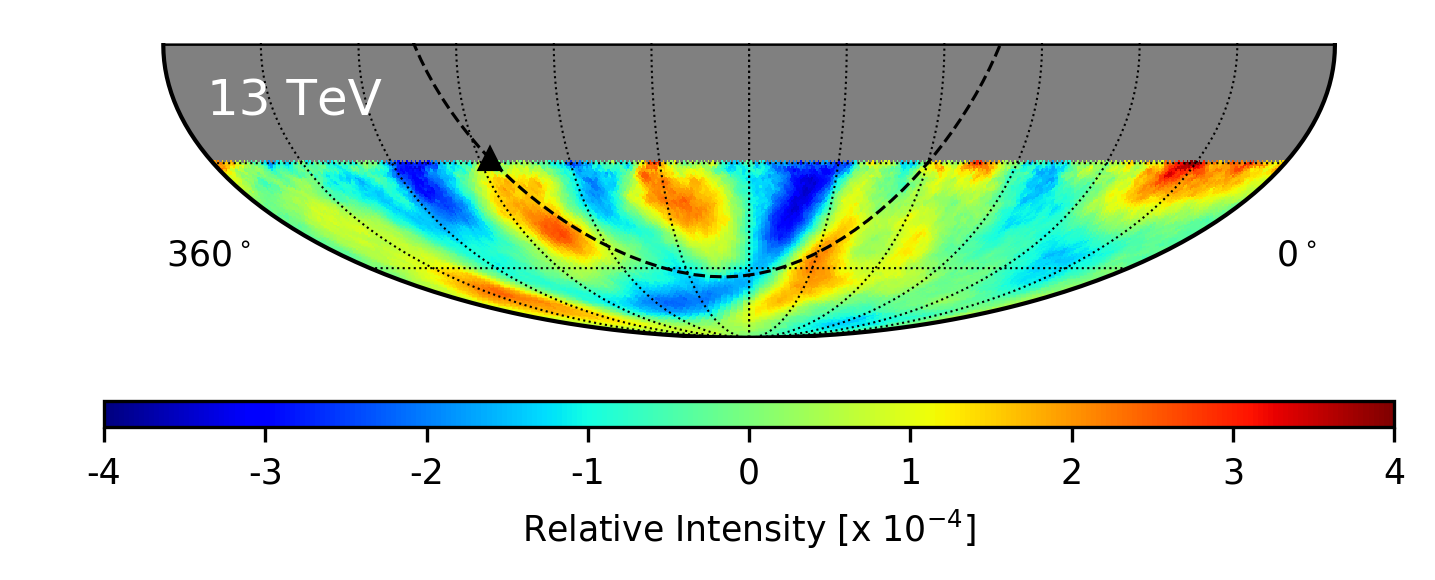}
  \includegraphics[width=0.49\textwidth]{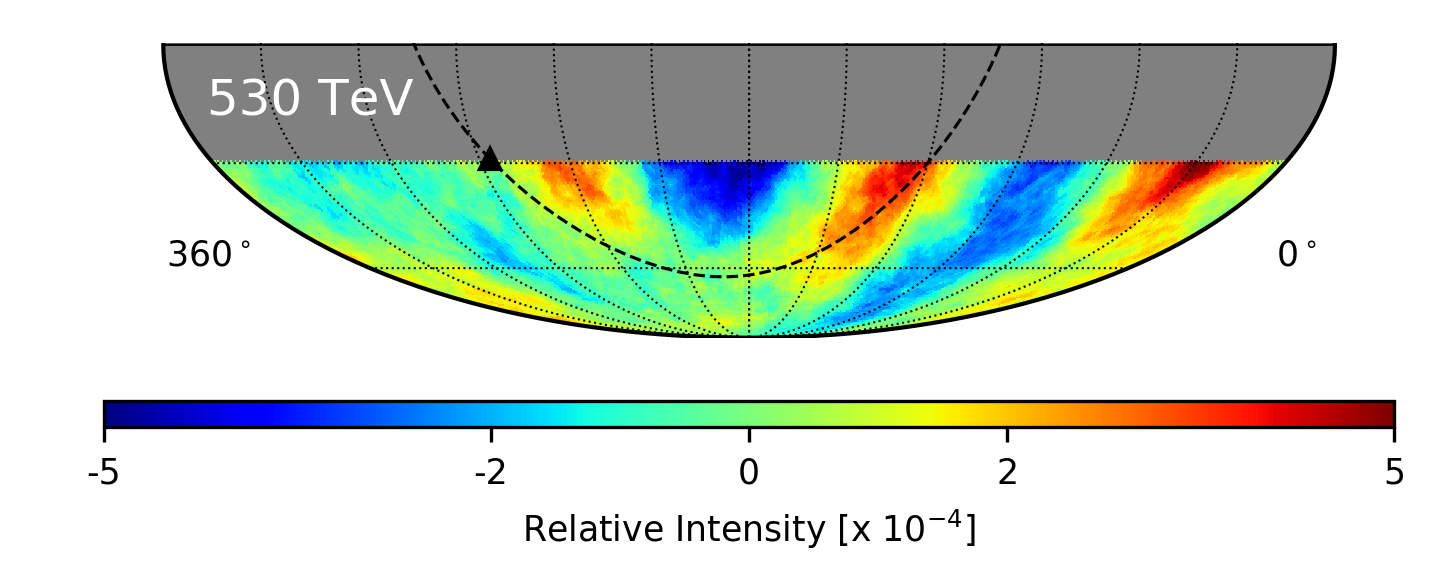}
  \includegraphics[width=0.49\textwidth]{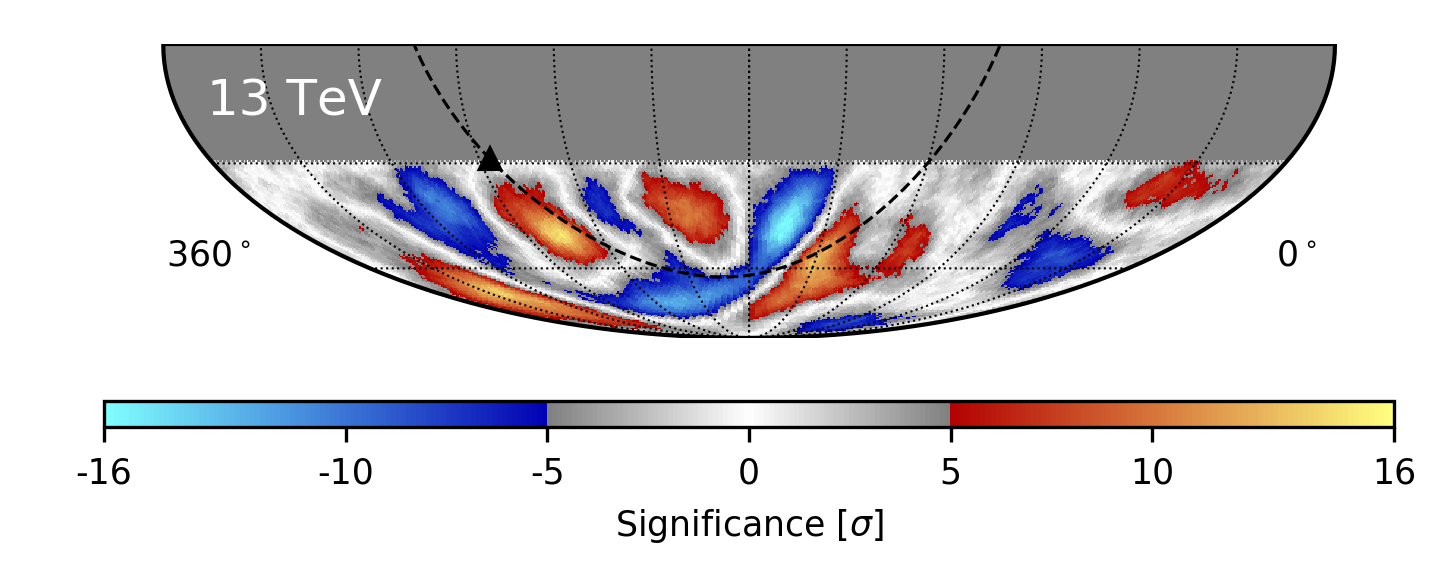}
  \includegraphics[width=0.49\textwidth]{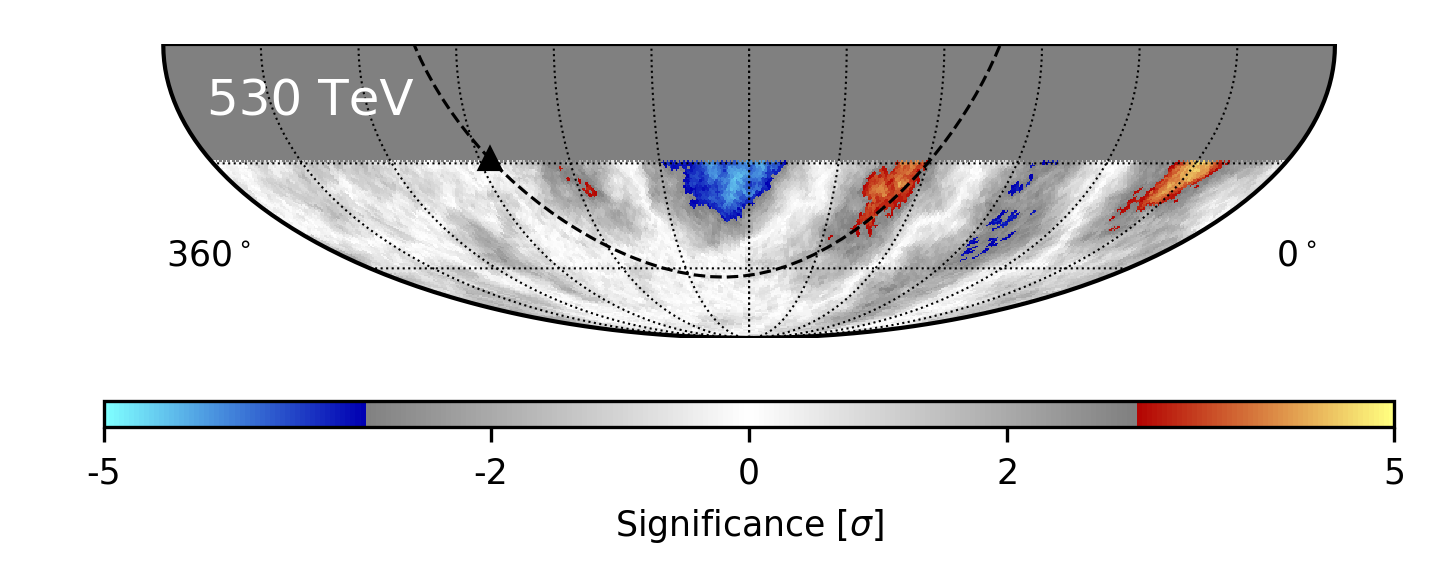}

  \caption{Relative intensity (\textit{top}) and statistical significance (\textit{bottom}) of the residual sky maps at
  low- (\textit{left}) and high-energy (\textit{right}) events ($E < 18$\,TeV and $> 320$\,TeV, respectively). The small-scale structure is made visible by subtracting the best-fit dipole and quadrupole components from the large-scale structure. All maps are shown in J2000 equatorial coordinates and use an angular smoothing radius of $5^{\circ}$ for the low-energy sample and $20^\circ$ for the high-energy one. The dashed line and triangle mark the galactic plane and center.}
  \label{fig:largesmall}
\end{figure*}

To emphasize small angular scale features in the maps (i.e., those represented by multipole moments with $\ell >$ 2), we produced residual sky maps by subtracting the best-fit dipole and quadrupole components on the sphere from the measured relative intensity maps. Because a major transition in the anisotropy occurs around 100\,TeV, we considered events with $E < 30$\,TeV and $E > 300$\,TeV separately. These energy ranges correspond to median energies of 13\,TeV and 530\,TeV, respectively, and have a fairly small overlap.
Figure~\ref{fig:largesmall} shows the residual anisotropy in relative intensity and the Li-Ma statistical significance for both low-energy and high-energy maps. In the figure, we use $5^{\circ}$ smoothing radius for the low-energy map and $20^{\circ}$ smoothing radius for the high-energy map.
The figure shows that the angular size of the residual anisotropy features remains roughly constant above and below 100 TeV, as also indicated by the relatively stable $\ell \simeq$ 6 -- 11 as a function of energy. However, the corresponding statistical significance decreases due to the steeply decreasing cosmic-ray energy spectrum.

\section{Systematics Studies} \label{sec:syst}

In this section, we describe the systematics studies we performed to validate the stability of our results. The interference between the cosmic-ray anisotropy in celestial coordinates and Earth's orbital Compton-Getting effect is described in Sec.~\ref{ssec:antisid}. We also assess the time stability of the observed anisotropy in Sec.~\ref{ssec:time}, and the effects of the FoV in the measurement of the angular power spectrum in Sec.~\ref{ssec:sysaps}.

\subsection{Interference with Earth's Orbital Motion} \label{ssec:antisid}

As the Earth revolves around the Sun, we observe an apparent excess of cosmic rays in the direction of the Earth's motion and a deficit in the opposite direction. 
Such an anisotropy, referred to as the \textit{Compton--Getting Effect}~\citep{Compton:1935,1968Ap&SS...2..431G} , explains the directional dependence caused by the movement of an observer with velocity $v$ through a flux of particles or photons. This effect can be observed in a solar time frame where the Sun remains at a fixed position and the Earth completes one revolution in 365.242 days. The resulting pattern of arrival directions is referred to as the ``solar dipole" and is a well-understood phenomenon used to verify the reliability and stability of an anisotropy analysis. We studied this effect extensively and we documented it in various publications such as~\cite{IceCube:2011oct,IceCube:2012feb,aartsen2016}.

At any time during the Earth's orbital motion, the solar dipole interferes with the observed cosmic-ray anisotropy in celestial coordinates (i.e., tied to the sidereal time frame). 
This interference is observable as an annual modulation of both the sidereal- and solar-frame anisotropies. If the measurements are performed in time intervals of one or more full years (i.e., with complete Earth revolutions), the annual interference across one or more orbital cycles cancels out. In the presence of data acquisition time gaps across the years, however, or in the case of the experiment's instrumental instabilities, the interference is not fully compensated by full orbital cycles, and both the cosmic-ray anisotropy in sidereal and solar frames are deformed by an amplitude and phase that depends on the data gaps and instabilities' occurrences in time (see~\cite{DiazVelez:2021zT} for more details). 

The interference between anisotropies originating in different time frames produces overlapping frequency sidebands. The sidereal day is 3.93 minutes \textit{shorter} than a solar day; to assess the impact of the orbital interference on the celestial signal, we can look at the anti-sidereal time frame, where a day is 3.93 minutes \textit{longer} than a solar day. The observation of a statistically significant antisidereal anisotropy can be used to either apply a correction to recover the \textit{actual} anisotropy (with a method developed by~\cite{Farley1954} and used by~\cite{guillian_2007}) or estimate a systematic error, as done in~\cite{aartsen2016}.

\begin{figure*}[t!]
  \centering
  \includegraphics[width=0.95\textwidth]{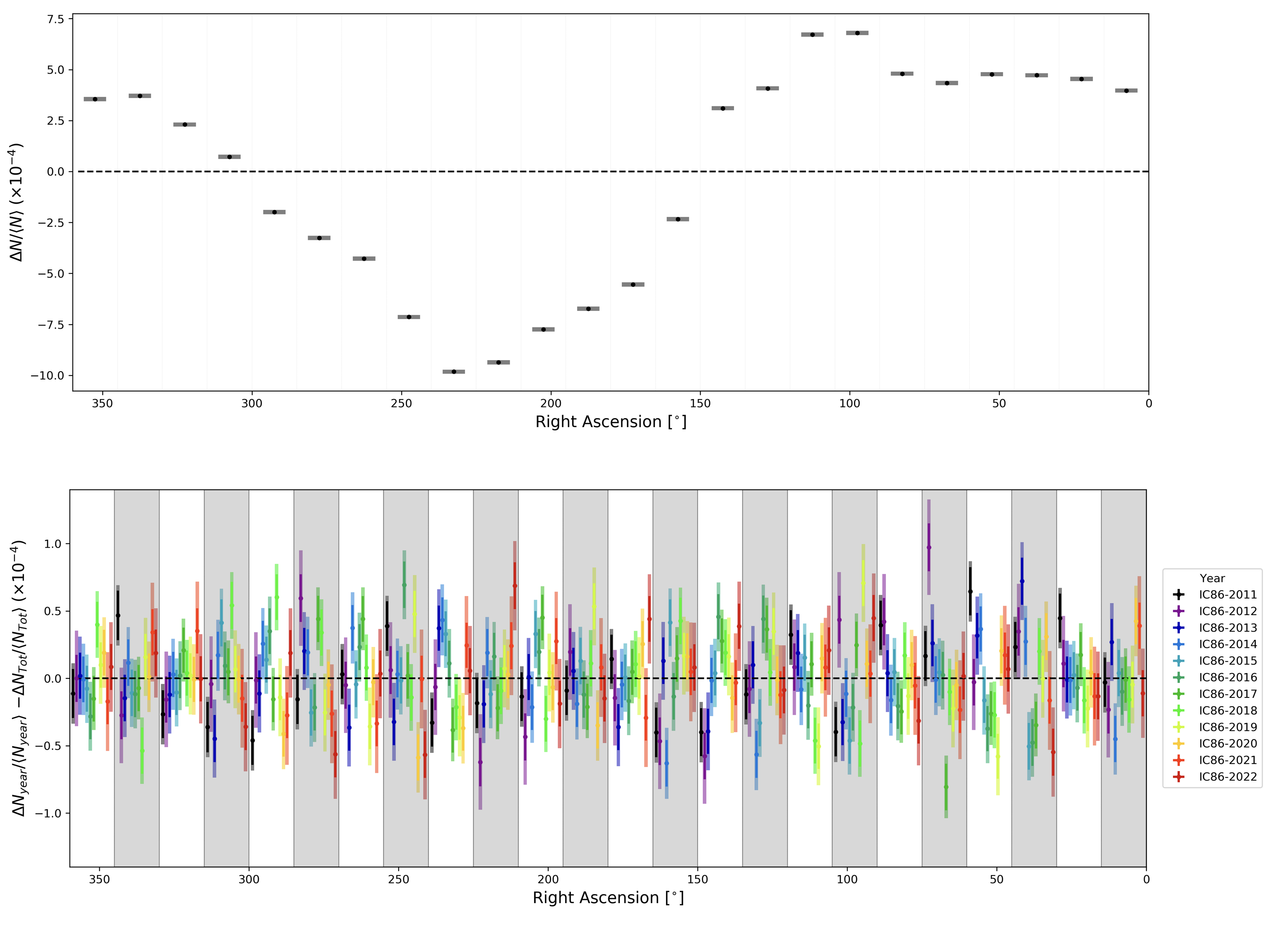}
  \caption{Top: One-dimensional projection of relative intensity as a function of RA (right ascension) for all data collected from May 13, 2011, to May 12, 2022. The grey band across each right ascension bin represents the average value for all years, and its thickness corresponds to the statistical uncertainty. Note that the statistical errors are approximately the same size as the symbols used in the graph. Bottom: Difference between each year's relative intensity and that of the entire data sample. The solid lines on each data point indicate the statistical errors, and the shaded boxes represent the systematic errors, which were calculated independently for each year using the anti-sidereal frame for each year.}
  \label{fig:time}
\end{figure*}

In this work, we produced cosmic-ray relative intensity sky maps in the anti-sidereal time frame, calculated the one-dimensional projection of relative intensity in the RA coordinates (see Sec.~\ref{ssec:dipole}), and used it to determine the RMS spread of the relative intensity at each RA bin for each year of our data sample. 
Figure~\ref{fig:time} shows that the systematic errors estimated with this method are of the same order as the statistical errors, a major improvement over the previous results published in~\cite{aartsen2016}. The dramatic reduction of the orbital interference systematic uncertainty is due to the data being split into calendar years. In past analyses, where we used data collected during IceCube's construction, we had to rely on the experiment's \textit{configuration years}, which were anywhere from 8 to 14 months long. Another reason for the small orbital interference systematic uncertainties stems from an absence of large data gaps due to the high duty cycle of the IceCube observatory over the last twelve years (see Table~\ref{table:data}). Histograms of time gaps between events showed no notable peaks beyond $10^{-2}$ seconds. Since the angular rotation of the Earth in $10^{-2}$\,s (($4.2 \times 10^{-5})^\circ$) is much less than the pixel size ($(0.84^\circ)^2$), any peaks on this scale and smaller should have no significant effect on our directional reconstruction.

\vskip 1.cm
\subsection{Anisotropy Stability Over Time} \label{ssec:time}

We investigated whether the cosmic-ray anisotropy is constant in time across the twelve years considered in the analysis. 
To do so, we looked at the one-dimensional projection of the relative intensity --- binned in RA --- for each calendar year. Figure~\ref{fig:time} shows the residual between each year's relative intensity from that of the entire data sample.
The solid error bars at each point are statistical uncertainties. The shaded boxes around each data point represent systematic uncertainties calculated from the anti-sidereal distribution for each year (see Sec.~\ref{ssec:antisid}). The effect of switching to calendar years is seen in the reduced size of the systematic uncertainties compared to previous analyses~\citep{aartsen2016} --- they are now comparable in size to statistical uncertainties. The black band at each RA bin corresponds to the relative intensity from the entire twelve-year sample.

To compare the year-to-year one-dimensional relative intensity distribution in RA, we performed a $\chi^2$ test. We found that each year is globally compatible with the twelve-year data within about $2\sigma$. However, this result does not exclude more localized time variations on the sky map from year to year. For this reason, we are currently conducting in-depth investigations to locate time variations in specific areas of the sky maps.

\subsection{Effects of Limited FoV on the Angular Power Spectrum} \label{ssec:sysaps}

The $a_{\ell m}$ multipole coefficients of the expansion (Eq. \ref{eq:sphHarmonics}) are given by
\begin{equation}\label{eq:exactalm}
a_{\ell m} = \int \delta I(\mathbf{\alpha, \delta})Y^*_{\ell m}(\mathbf{\alpha, \delta}) d\Omega \, ,
\end{equation}
which, in the case of a pixelized sky map with limited FoV, becomes
\begin{equation}\label{eq:alm}
  \hat{a}_{\ell m} \simeq \Omega_p\sum_{i\in \mathrm{FoV}}\delta I(\mathbf{u}_i)Y^*_{\ell m}(\mathbf{u}_i) \, , 
\end{equation}
summed over the pixels in the FoV, where $\Omega_p$ is the solid angle observed by each pixel i, $\mathbf{u}_i = (\alpha_i, \delta_i)$ is the pointing vector associated with the ith pixel. As a result of the time--scrambling method, the relative intensity $\delta I$ is a function of $\alpha$, centered around zero at any declination, while its amplitude depends on $\delta$ from the spherical harmonics functions. 
Therefore, $\delta I (\alpha, \delta) = f(\delta)g(\alpha)$ and

\begin{align}
  \label{eq:alm0}
  \begin{split}
    \hspace*{-0.6cm}
    \hat{a}_{\ell 0} &= \int^{\delta_\mathrm{max}}_{\delta_\mathrm{min}}   \int^{2\pi}_0 Y_{\ell 0}(\delta)\delta I(\alpha,\delta)\, d\alpha\, d\cos \delta \\
    &= \underbrace{\int^{\delta_\mathrm{max}}_{\delta_\mathrm{min}}Y_{\ell 0}(\delta)f(\delta)\, d\cos \delta}_{\neq\, 0}
                   \underbrace{\int^{2\pi}_0 g(\alpha)\, d\alpha}_{=\, 0} \\
    &= 0
  \end{split}
\end{align}

The fact that $\hat{a}_{\ell 0} \sim 10^{-10}$ from the sum in Eq. (\ref{eq:alm})  with $m=0$ are not identically zero for the data in this analysis is the result of Poisson fluctuations in the data and background and numerical errors from the pixelization of the sky maps. 

As mentioned in Sec.~\ref{ssec:ang} and discussed in~\cite{SOMMERS2001271} and \cite{Abeysekara_2019}, the limited FoV of the observations introduces correlations between different $\ell$ modes which decrease for larger values of $\ell$. 
\begin{figure*}[ht]
  \centering
    \includegraphics[width=.49\textwidth]{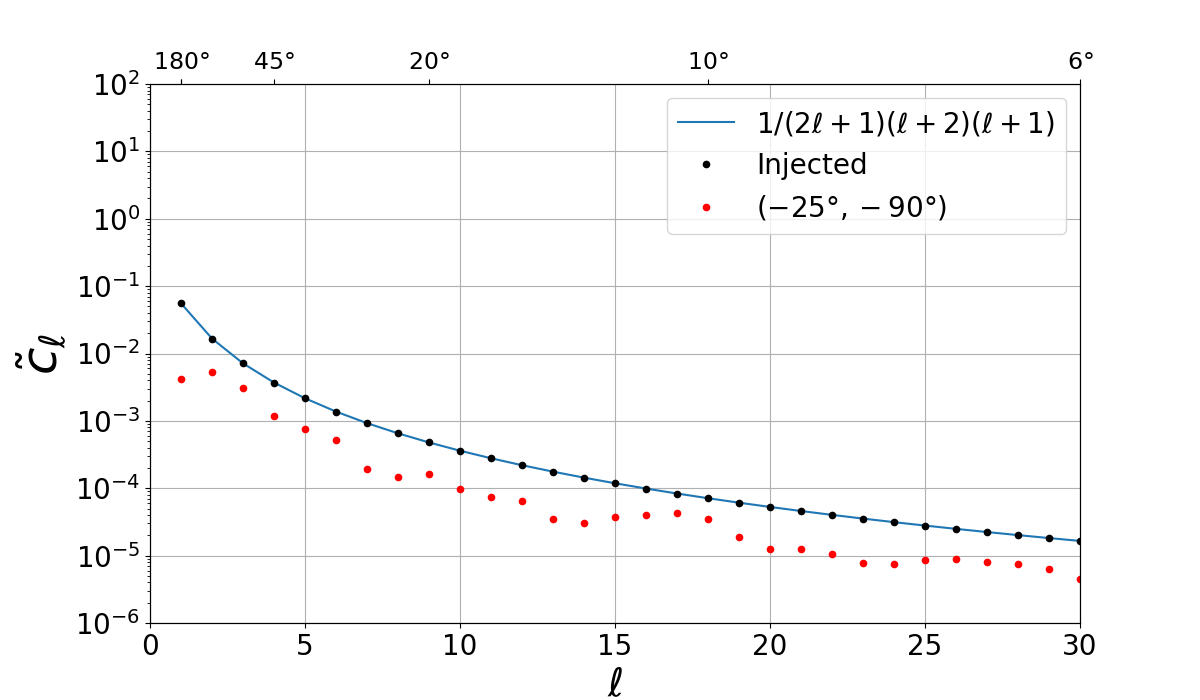}
    \includegraphics[width=.49\textwidth]{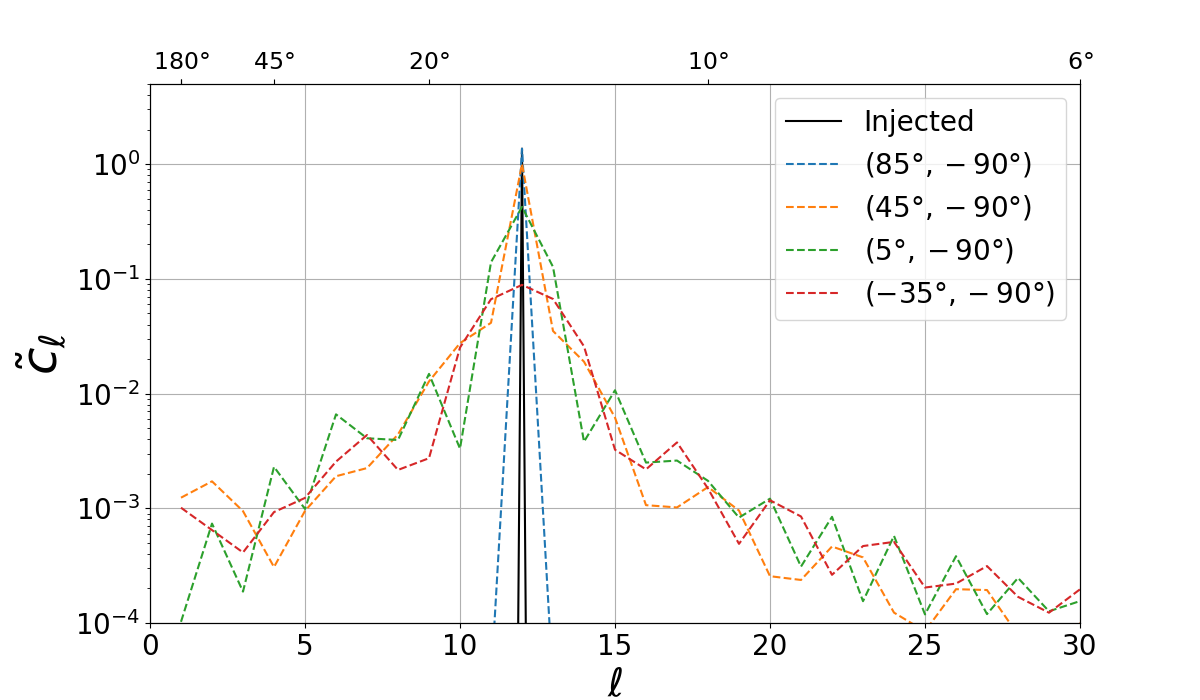}
    \caption{Left: Angular pseudo-power spectrum obtained from injected angular power distribution after masking two-thirds of the sky.
    The resulting spectrum globally decreases due to the limited FoV. In addition, the power of different multipoles is reduced unevenly for different values of $\ell$. Right: Angular pseudo-power spectrum obtained from injected Kronecker delta function. The declination range in the legend indicates the unmasked region.}
    \label{fig:toyaps}
\end{figure*}
To better understand the role of the limited FoV on the angular pseudo-power spectrum, we generated random full-sky maps using the \emph{synfast} function from \emph{healpy}
based on an arbitrary angular power spectrum given by 
\begin{equation}\label{eq:toycl}
    C_\ell \propto \frac{1}{(2\ell+1)(\ell+2)(\ell+1)}\, , 
\end{equation}
introduced by~\cite{Ahlers:2014jan}, that roughly mimics the observed angular power spectrum from the cosmic-ray arrival direction distribution. The generated maps conserve the $C_\ell$ power but are randomly distributed across the $a_{\ell m}$ values. We then calculate new $\tilde{C}_\ell$ values after masking the northern sky outside the FoV of IceCube. Since the multipole modes are randomly distributed across the sky, the resulting angular pseudo-power spectrum decreases globally due to the limited FoV. This is illustrated on the left of Fig.~\ref{fig:toyaps}. Also, the power of a given multipole term $\ell$ is spread across neighboring $\ell$ values due to the correlation between $\ell$ modes resulting from the limited FoV. The right side of Fig.~\ref{fig:toyaps} displays the power-smearing of an injected Kronecker delta function ($\delta_{\ell=12}$) from the limited FoV.

These two effects are responsible for an uneven power reduction vs. $\ell$, as illustrated on the left of Fig.~\ref{fig:toyaps}, which qualitatively resembles the measured pseudo-power spectrum of Fig.~\ref{fig:ebinned_aps}.
To properly describe the angular power distribution, performing an all-sky analysis similar to what was done by \cite{Abeysekara_2018} is important. 

\section{Summary} \label{sec:summary}

This work provides the latest analysis of the distribution of cosmic-ray arrival directions in the Southern Hemisphere. The study is based on the largest data sample ever collected by the IceCube in-ice array, which consists of $7.92 \times 10^{11}$ cosmic ray-induced muon events collected over a 12-year period. The study covers more than a full period length of a solar cycle.
We took advantage of the completed configuration of the IceCube instrumentation in that the experiment's response function is constant over the entire time period considered. An improved Monte Carlo simulation description of our events now provides a stable and reliable cosmic-ray muon detection response and energy estimation across the entire sample. The resulting stability of our data reduces systematic uncertainties below statistical fluctuations compared to previous publications. The larger data volume enables improving the statistical significance of measurements up to a few PeV and down to about 6$^{\circ}$ angular scale. 

We produced sky maps in relative intensity and Li-Ma statistical significance of the cosmic-ray arrival direction from a median energy of 13\,TeV to over 5\,PeV. We confirm the observation of a change in the angular structure of the cosmic-ray anisotropy in this energy range, mainly occurring within 100--300\,TeV. 
The average angular scale of features in Figs.~\ref{fig:eplots} and~\ref{fig:eplotssig} appears to shrink with increasing energy. On the other hand, the small angular scale features highlighted in Fig.~\ref{fig:largesmall} appear to be similarly sized both at low and high energies (13 and 530\,TeV, respectively).
This observation is quantified by the corresponding angular pseudo-power spectra in Figs.~\ref{fig:ebinned_aps} and~\ref{fig:apse}, where we observe a variation in the angular pseudo-power spectrum as a function of energy, hinting at relatively reduced large-scale features at high energy compared to those of medium and small scales. 

In particular, Fig.~\ref{fig:apse} shows that the decrease of the angular pseudo-power $\tilde{C}_\ell$ for structures with $\ell \leq 3$ appear to follow a similar trend as the amplitude of the dipole component in Fig.~\ref{fig:dipole}, while the power of smaller structures of $\ell \geq 6$ appears to remain relatively constant. Several authors have suggested the existence of two different mechanisms responsible for the observed angular scale features. The larger $\ell$ values seem to be consistent with pitch-angle scattering effects on magnetic turbulence within the mean free path~\citep{PhysRevLett.109.071101, Ahlers:2014jan, Ahlers:2015dwa, kuhlen2021cosmic, Kuhlen_2022, zhang2024smallscale}, while the lower $\ell$ may be associated with large-scale diffusive processes (over many mean free paths) across the interstellar medium~\citep{erlykin_2006,Blasi:2011fi,Ptuskin:2012dec,pohl_2013,Sveshnikova:2013dec,savchenko_2015,ahlers_2016,Giacinti_2017}.

We often use the dipole component of the observed anisotropy as a proxy for cosmic-ray diffusive models in the ISM. However, the abovementioned studies highlight the need to carefully measure the cosmic-ray arrival direction distribution at small scales (i.e., with high-$\ell$ multipole moments) as the angular power spectrum reflects the turbulence present in interstellar space.
To determine the dipole component, we updated our measurement by fitting a two-dimensional arrival direction distribution sky map that takes into account the limited FoV of the IceCube in-ice array (see Sec.~\ref{ssec:dipole}). In Fig.~\ref{fig:dipole}, we have presented the amplitude and phase of the dipole component and compared our results with those of other experiments. We have also presented the angular pseudo-power spectrum to show the energy dependence of the different multipole modes and highlighted the importance of full-sky cosmic-ray anisotropy observations.
It is important to note that all reported observations are only sensitive to the horizontal components of the anisotropy, as the data-driven estimation of the reference maps absorbs the equatorial north-south component. 
This analysis shows that each year's one-dimensional distribution of relative intensity as a function of RA is globally compatible with the twelve-year data sample within about 2$\sigma$. However, this does not exclude more localized time variations on the sky map from year to year.

\section{Outlook} \label{sec:outlook}

Current efforts to further advance our understanding of cosmic rays' properties include a follow-up study with the IceTop surface array on anisotropy observations. The IceTop subdetector, which is not used in this analysis, is sensitive to a per-mille anisotropy across various energy bands around the PeV region. This energy range is particularly noteworthy, displaying distinct structures in the chemical composition and the cosmic-ray energy spectrum. Examining IceTop data in the 1-10\,PeV region enables a direct comparison of energy-dependent cosmic-ray anisotropy with the IceCube in-ice array, and it also allows for a comprehensive exploration of potential correlations between arrival direction anisotropy, energy spectrum, and chemical composition of the cosmic-ray flux. This, in turn, will allow us to study the effect of the increasing magnetic rigidity of the primary particles on the strength and angular distribution of the anisotropy. 

A crucial step in advancing our understanding of cosmic-ray anisotropy and overcoming the limitations of single-detector observations is to extend our observational scope from the Southern Hemisphere to full-sky observations with multiple detectors in both hemispheres. The sky maps from experiments in the Northern Hemisphere combined with the IceCube in-ice array and IceTop surface array observations would enable us to mitigate biases discussed in this paper attributed to the exposure to a limited portion of the sky. For energies surpassing 1\,PeV, a full sky map allows a better understanding of cosmic rays' pitch-angle distribution beyond the heliosphere's influence.

We also plan to extend our investigations of cosmic-ray anisotropy to the time domain and search for possible localized variations of the cosmic-ray flux on the sky map. By analyzing the time variations in the distribution of cosmic ray arrival directions, we hope to identify any potential correlation with heliospheric modulations at the boundary with the interstellar medium. Moreover, we will explore whether the features in the arrival direction distribution of cosmic rays are associated with modulations in their energy spectrum.

These observations, especially if combined with other experiments, may increase our understanding of the phenomenology responsible for the anisotropy. Specifically, these observations will help us study the influence of magnetic field structures in our local interstellar medium, the propagation properties of interstellar turbulent plasma, and the diffusion propagation across the Milky Way.

Future advancements in instrumentation, such as IceCube-Gen2 and the Southern Wide-field Gamma-ray Observatory (SWGO), are expected to enhance our observations even further. IceCube-Gen2 is designed to have a broader aperture, which will enable surface-to-in-ice coincidence studies, consequently improving energy and mass sensitivity in the PeV energy range~\citep{Aartsen_2021, 2023EPJWC.28301001S, IceCube-Gen2:2023}. In addition, the SWGO observatory in the Southern Hemisphere will provide modern instrumentation with improved energy and mass composition sensitivity and significantly contribute to cosmic-ray physics, including anisotropy measurements~\citep{Taylor:20210}.

\section*{Acknowledgements}
\input{ack}

\bibliography{tenyear_cra}
\bibliographystyle{aasjournal}



\end{document}

%% file: authors_i3.tex
\affiliation{III. Physikalisches Institut, RWTH Aachen University, D-52056 Aachen, Germany}
\affiliation{Department of Physics, University of Adelaide, Adelaide, 5005, Australia}
\affiliation{Dept. of Physics and Astronomy, University of Alaska Anchorage, 3211 Providence Dr., Anchorage, AK 99508, USA}
\affiliation{Dept. of Physics, University of Texas at Arlington, 502 Yates St., Science Hall Rm 108, Box 19059, Arlington, TX 76019, USA}
\affiliation{School of Physics and Center for Relativistic Astrophysics, Georgia Institute of Technology, Atlanta, GA 30332, USA}
\affiliation{Dept. of Physics, Southern University, Baton Rouge, LA 70813, USA}
\affiliation{Dept. of Physics, University of California, Berkeley, CA 94720, USA}
\affiliation{Lawrence Berkeley National Laboratory, Berkeley, CA 94720, USA}
\affiliation{Institut f{\"u}r Physik, Humboldt-Universit{\"a}t zu Berlin, D-12489 Berlin, Germany}
\affiliation{Fakult{\"a}t f{\"u}r Physik {\&} Astronomie, Ruhr-Universit{\"a}t Bochum, D-44780 Bochum, Germany}
\affiliation{Universit{\'e} Libre de Bruxelles, Science Faculty CP230, B-1050 Brussels, Belgium}
\affiliation{Vrije Universiteit Brussel (VUB), Dienst ELEM, B-1050 Brussels, Belgium}
\affiliation{Dept. of Physics, Simon Fraser University, Burnaby, BC V5A 1S6, Canada}
\affiliation{Department of Physics and Laboratory for Particle Physics and Cosmology, Harvard University, Cambridge, MA 02138, USA}
\affiliation{Dept. of Physics, Massachusetts Institute of Technology, Cambridge, MA 02139, USA}
\affiliation{Dept. of Physics and The International Center for Hadron Astrophysics, Chiba University, Chiba 263-8522, Japan}
\affiliation{Department of Physics, Loyola University Chicago, Chicago, IL 60660, USA}
\affiliation{Dept. of Physics and Astronomy, University of Canterbury, Private Bag 4800, Christchurch, New Zealand}
\affiliation{Dept. of Physics, University of Maryland, College Park, MD 20742, USA}
\affiliation{Dept. of Astronomy, Ohio State University, Columbus, OH 43210, USA}
\affiliation{Dept. of Physics and Center for Cosmology and Astro-Particle Physics, Ohio State University, Columbus, OH 43210, USA}
\affiliation{Niels Bohr Institute, University of Copenhagen, DK-2100 Copenhagen, Denmark}
\affiliation{Dept. of Physics, TU Dortmund University, D-44221 Dortmund, Germany}
\affiliation{Dept. of Physics and Astronomy, Michigan State University, East Lansing, MI 48824, USA}
\affiliation{Dept. of Physics, University of Alberta, Edmonton, Alberta, T6G 2E1, Canada}
\affiliation{Erlangen Centre for Astroparticle Physics, Friedrich-Alexander-Universit{\"a}t Erlangen-N{\"u}rnberg, D-91058 Erlangen, Germany}
\affiliation{Physik-department, Technische Universit{\"a}t M{\"u}nchen, D-85748 Garching, Germany}
\affiliation{D{\'e}partement de physique nucl{\'e}aire et corpusculaire, Universit{\'e} de Gen{\`e}ve, CH-1211 Gen{\`e}ve, Switzerland}
\affiliation{Dept. of Physics and Astronomy, University of Gent, B-9000 Gent, Belgium}
\affiliation{Dept. of Physics and Astronomy, University of California, Irvine, CA 92697, USA}
\affiliation{Karlsruhe Institute of Technology, Institute for Astroparticle Physics, D-76021 Karlsruhe, Germany}
\affiliation{Karlsruhe Institute of Technology, Institute of Experimental Particle Physics, D-76021 Karlsruhe, Germany}
\affiliation{Dept. of Physics, Engineering Physics, and Astronomy, Queen's University, Kingston, ON K7L 3N6, Canada}
\affiliation{Department of Physics {\&} Astronomy, University of Nevada, Las Vegas, NV 89154, USA}
\affiliation{Nevada Center for Astrophysics, University of Nevada, Las Vegas, NV 89154, USA}
\affiliation{Dept. of Physics and Astronomy, University of Kansas, Lawrence, KS 66045, USA}
\affiliation{Centre for Cosmology, Particle Physics and Phenomenology - CP3, Universit{\'e} catholique de Louvain, Louvain-la-Neuve, Belgium}
\affiliation{Department of Physics, Mercer University, Macon, GA 31207-0001, USA}
\affiliation{Dept. of Astronomy, University of Wisconsin{\textemdash}Madison, Madison, WI 53706, USA}
\affiliation{Dept. of Physics and Wisconsin IceCube Particle Astrophysics Center, University of Wisconsin{\textemdash}Madison, Madison, WI 53706, USA}
\affiliation{Institute of Physics, University of Mainz, Staudinger Weg 7, D-55099 Mainz, Germany}
\affiliation{Department of Physics, Marquette University, Milwaukee, WI 53201, USA}
\affiliation{Institut f{\"u}r Kernphysik, Westf{\"a}lische Wilhelms-Universit{\"a}t M{\"u}nster, D-48149 M{\"u}nster, Germany}
\affiliation{Bartol Research Institute and Dept. of Physics and Astronomy, University of Delaware, Newark, DE 19716, USA}
\affiliation{Dept. of Physics, Yale University, New Haven, CT 06520, USA}
\affiliation{Columbia Astrophysics and Nevis Laboratories, Columbia University, New York, NY 10027, USA}
\affiliation{Dept. of Physics, University of Oxford, Parks Road, Oxford OX1 3PU, United Kingdom}
\affiliation{Dipartimento di Fisica e Astronomia Galileo Galilei, Universit{\`a} Degli Studi di Padova, I-35122 Padova PD, Italy}
\affiliation{Dept. of Physics, Drexel University, 3141 Chestnut Street, Philadelphia, PA 19104, USA}
\affiliation{Physics Department, South Dakota School of Mines and Technology, Rapid City, SD 57701, USA}
\affiliation{Dept. of Physics, University of Wisconsin, River Falls, WI 54022, USA}
\affiliation{Dept. of Physics and Astronomy, University of Rochester, Rochester, NY 14627, USA}
\affiliation{Department of Physics and Astronomy, University of Utah, Salt Lake City, UT 84112, USA}
\affiliation{Dept. of Physics, Chung-Ang University, Seoul 06974, Republic of Korea}
\affiliation{Oskar Klein Centre and Dept. of Physics, Stockholm University, SE-10691 Stockholm, Sweden}
\affiliation{Dept. of Physics and Astronomy, Stony Brook University, Stony Brook, NY 11794-3800, USA}
\affiliation{Dept. of Physics, Sungkyunkwan University, Suwon 16419, Republic of Korea}
\affiliation{Institute of Basic Science, Sungkyunkwan University, Suwon 16419, Republic of Korea}
\affiliation{Institute of Physics, Academia Sinica, Taipei, 11529, Taiwan}
\affiliation{Dept. of Physics and Astronomy, University of Alabama, Tuscaloosa, AL 35487, USA}
\affiliation{Dept. of Astronomy and Astrophysics, Pennsylvania State University, University Park, PA 16802, USA}
\affiliation{Dept. of Physics, Pennsylvania State University, University Park, PA 16802, USA}
\affiliation{Dept. of Physics and Astronomy, Uppsala University, Box 516, SE-75120 Uppsala, Sweden}
\affiliation{Dept. of Physics, University of Wuppertal, D-42119 Wuppertal, Germany}
\affiliation{Deutsches Elektronen-Synchrotron DESY, Platanenallee 6, D-15738 Zeuthen, Germany}

\author[0000-0001-6141-4205]{R. Abbasi}
\affiliation{Department of Physics, Loyola University Chicago, Chicago, IL 60660, USA}

\author[0000-0001-8952-588X]{M. Ackermann}
\affiliation{Deutsches Elektronen-Synchrotron DESY, Platanenallee 6, D-15738 Zeuthen, Germany}

\author{J. Adams}
\affiliation{Dept. of Physics and Astronomy, University of Canterbury, Private Bag 4800, Christchurch, New Zealand}

\author[0000-0002-9714-8866]{S. K. Agarwalla}
\altaffiliation{also at Institute of Physics, Sachivalaya Marg, Sainik School Post, Bhubaneswar 751005, India}
\affiliation{Dept. of Physics and Wisconsin IceCube Particle Astrophysics Center, University of Wisconsin{\textemdash}Madison, Madison, WI 53706, USA}

\author{T. Aguado}
\affiliation{Department of Physics, Loyola University Chicago, Chicago, IL 60660, USA}

\author[0000-0003-2252-9514]{J. A. Aguilar}
\affiliation{Universit{\'e} Libre de Bruxelles, Science Faculty CP230, B-1050 Brussels, Belgium}

\author[0000-0003-0709-5631]{M. Ahlers}
\affiliation{Niels Bohr Institute, University of Copenhagen, DK-2100 Copenhagen, Denmark}

\author[0000-0002-9534-9189]{J.M. Alameddine}
\affiliation{Dept. of Physics, TU Dortmund University, D-44221 Dortmund, Germany}

\author{N. M. Amin}
\affiliation{Bartol Research Institute and Dept. of Physics and Astronomy, University of Delaware, Newark, DE 19716, USA}

\author[0000-0001-9394-0007]{K. Andeen}
\affiliation{Department of Physics, Marquette University, Milwaukee, WI 53201, USA}

% ...

\date{\today}

\collaboration{437}{IceCube Collaboration}

%% file: ack.tex
The IceCube collaboration acknowledges the significant contributions to this manuscript from Loyola University Chicago, Mercer University, and the University of Wisconsin -- Madison.
The authors gratefully acknowledge the support from the following agencies and institutions:
USA {\textendash} U.S. National Science Foundation-Office of Polar Programs,
U.S. National Science Foundation-Physics Division,
U.S. National Science Foundation-EPSCoR,
U.S. National Science Foundation REU,
U.S. National Science Foundation-Office of Advanced Cyberinfrastructure,
Wisconsin Alumni Research Foundation,
Center for High Throughput Computing (CHTC) at the University of Wisconsin{\textendash}Madison,
Open Science Grid (OSG),
Partnership to Advance Throughput Computing (PATh),
Advanced Cyberinfrastructure Coordination Ecosystem: Services {\&} Support (ACCESS),
Frontera computing project at the Texas Advanced Computing Center,
U.S. Department of Energy-National Energy Research Scientific Computing Center,
Particle astrophysics research computing center at the University of Maryland,
Institute for Cyber-Enabled Research at Michigan State University,
Astroparticle physics computational facility at Marquette University,
NVIDIA Corporation,
and Google Cloud Platform;
Belgium {\textendash} Funds for Scientific Research (FRS-FNRS and FWO),
FWO Odysseus and Big Science programmes,
and Belgian Federal Science Policy Office (Belspo);
Germany {\textendash} Bundesministerium f{\"u}r Bildung und Forschung (BMBF),
Deutsche Forschungsgemeinschaft (DFG),
Helmholtz Alliance for Astroparticle Physics (HAP),
Initiative and Networking Fund of the Helmholtz Association,
Deutsches Elektronen Synchrotron (DESY),
and High Performance Computing cluster of the RWTH Aachen;
Sweden {\textendash} Swedish Research Council,
Swedish Polar Research Secretariat,
Swedish National Infrastructure for Computing (SNIC),
and Knut and Alice Wallenberg Foundation;
European Union {\textendash} EGI Advanced Computing for research;
Australia {\textendash} Australian Research Council;
Canada {\textendash} Natural Sciences and Engineering Research Council of Canada,
Calcul Qu{\'e}bec, Compute Ontario, Canada Foundation for Innovation, WestGrid, and Digital Research Alliance of Canada;
Denmark {\textendash} Villum Fonden, Carlsberg Foundation, and European Commission;
New Zealand {\textendash} Marsden Fund;
Japan {\textendash} Japan Society for Promotion of Science (JSPS)
and Institute for Global Prominent Research (IGPR) of Chiba University;
Korea {\textendash} National Research Foundation of Korea (NRF);
Switzerland {\textendash} Swiss National Science Foundation (SNSF).